\documentclass[preprint]{aastex}
\usepackage{psfig,emulateapj5}

\newcommand{\kms}{km~s$^{-1}$}

\begin{document}
\slugcomment{To appear in the Astronomical Journal}
\title{Kinematic Constraints on Evolutionary Scenarios for Blue Compact Dwarf Galaxies : I. Neutral Gas Dynamics}
\author{Liese van~Zee}
\affil{Herzberg Institute of Astrophysics, National Research Council of Canada \\ 5071 W. 
Saanich Road, Victoria, BC V9E 2E7 Canada}
\email{Liese.vanZee@hia.nrc.ca}
\author{John J. Salzer\footnote{NSF Presidential Faculty Fellow}}
\affil{Astronomy Department, Wesleyan University, Middletown, CT 06459--0123}
\email{slaz@parcha.astro.wesleyan.edu}
\and
\author{Evan D. Skillman}
\affil{Astronomy Department, University of Minnesota, 116 Church St. SE}
\affil{Minneapolis, MN 55455}
\email{skillman@astro.spa.umn.edu}
\begin{abstract}
We present the results of high spatial resolution
HI synthesis observations of six blue
compact dwarf (BCD) galaxies.  Optically, the selected galaxies  
have smooth, symmetric isophotes, and thus
are the most likely of the BCD class to fade into an object
morphologically similar
to a dwarf elliptical when the current starburst ends.
The neutral gas in all six galaxies  appears to be
rotationally supported, however, indicating that true
morphological transformation from a BCD to a dE will 
require significant loss of angular momentum.  Based on the
observed neutral gas dynamics of these and other BCDs, 
it is unlikely that present--day BCDs will evolve directly into 
dwarf ellipticals after a starburst phase.  We discuss alternative
evolutionary scenarios for BCDs and place them within the larger context
of galaxy formation and evolution models.
 
In general, BCDs appear to have steeper rotation curves than
similar luminosity low surface brightness dwarf galaxies.
BCDs have centrally concentrated mass distributions
(stars, gas, and dark matter) and have lower specific angular
momenta.  Based on disk instability analyses, steeply rising rotation 
curves result in higher threshold densities for the onset of star formation.
These results suggest that angular momenta may play
a crucial role in the morphological evolution of low mass galaxies:
galaxies with low angular momenta will be able to collapse into
small, compact galaxies while galaxies with high angular momenta will
be more diffuse systems.  Further, because the star formation threshold
density is higher in  low angular momenta systems, star formation will be delayed
until an extremely high surface density is reached.  Thus, angular momentum
may be the fundamental parameter which determines if a low mass galaxy
will have centrally concentrated stellar and gaseous distributions,
and be more susceptible to a burst mode of star formation.
\end{abstract}


\section{Introduction}
\label{sec:intro}
Blue compact dwarf (BCD) galaxies pose an interesting puzzle for
standard galaxy evolution models. Their extremely low
gas--phase metallicities and blue colors suggest that
these galaxies are relatively unevolved systems, perhaps
undergoing their first bursts of star formation at the
present epoch \citep[e.g.,][]{SS70,IT99,TIF99}.
Recently, however, deep optical and infrared imaging
of BCDs have shown that the majority have extensive low surface 
brightness stellar halos, indicating that most have an underlying older
stellar population in addition to bright young stars from the
present starburst \citep[e.g.,][]{J94,PLFT96,PLTF96}.
 Thus, despite their low luminosity, low metallicity nature, most BCDs
are not young systems; rather, the burst phase in BCDs 
is a transitory event, and perhaps cyclical in nature.

This immediately raises the question: where are the quiescent--phase 
BCDs today?  Are they included as part of the large, diverse, class of dwarf
irregulars?  Do BCDs use up their gas and fade into dwarf ellipticals?
Or, do they fade beyond detectability, and thus lie hidden 
and ignored in the vastness of space?  The possibility of morphological
evolution among the dwarf galaxy classifications (dE, dI, BCD,
in order of increasing star formation activity) has been suggested
many times over the last several decades \citep[e.g.,][]{LT86,SWS87,DP88,HHSL89,DH91,J94,FB94,PLFT96},
with a long list of positive and negative attributes for each of the various possible
transformations.  One of the popular evolutionary scenarios
is that the starburst phase uses up \citep[or ``blows--out,''][]{DS86} 
the interstellar medium.  As the
young stars fade, the system will then evolve into a gas--poor, red, dwarf
elliptical.  Alternatively, if the gas is not entirely consumed,
the system could continue to form stars at a much lower rate,
and thus be classified as a gas--rich dwarf irregular.  

The possibility of evolutionary connections between BCDs, dIs, and 
dEs has largely been based on the remarkable similarity in their stellar 
distributions.  While exceptions exist, the majority of these low mass 
objects are well--fit by exponential disks; furthermore, in a plot of 
magnitude vs.\ surface brightness, the dE class is distinct 
from the giant elliptical class, but overlaps with the dIs
\citep[e.g.,][]{FB94}.
The similarity in stellar distributions (but not current stellar populations
or gas content) suggested that dEs could be faded dIs, or in some cases, such as the
nucleated dwarf ellipticals, faded BCDs.  

Recently, detailed kinematic studies of
BCDs, dEs, and dIs has shed new light on the interrelationship between
these low mass galaxies.
High resolution HI kinematic studies of a number of
BCDs indicate that these are rotation dominated systems 
\citep[e.g.,][]{WBDK97,vSS98,vWHS98}.
In contrast, stellar kinematic observations of dEs find little to no
evidence of rotational support \citep{BN90,BPN91}; if the
gas and stars are kinematically coupled in BCDs, these results
suggest that BCDs cannot evolve into dEs because of angular
momentum conservation considerations.  
However, the BCD classification covers a large range of
morphological types of ``host'' galaxies  \citep[e.g.,][]{LT85},
and the previous samples did not include systems that were 
necessarily likely to evolve into dEs.  Thus, finding rotation in 
those BCDs does not preclude this scenario for all types 
of BCDs.  In order to test directly the feasibility of the popular
BCD to dE evolutionary scenario, we have obtained HI synthesis observations
of a small sample of BCDs which appeared
to be ``hosted'' by dwarf elliptical--like systems; that is, this
sample contains the systems
that are {\em most likely} to be morphologically similar to dEs when
their starburst fades.

In this paper, we present the results of high spatial resolution
HI synthesis observations of six BCDs. We discuss the sample 
selection and observational procedures in
Section 2.  The HI distributions and kinematics of the six
galaxies are presented in Section 3. 
A discussion of the neutral gas dynamics and its implications for BCD evolutionary
scenarios is presented in Section 4; Section 5 contains a
brief summary of the conclusions.

\section{Data Acquisition and Analysis}
\label{sec:data}
Multiconfiguration HI observations of six BCDs were obtained with
the Very Large Array\footnote{The Very Large Array
is a facility of the National Radio Astronomy Observatory.
The National Radio 
Astronomy Observatory is a facility of the National Science Foundation,
operated under a cooperative agreement by Associated Universities Inc.}. 
In this section, the sample selection and procedures for
observation and data reduction are discussed.

\subsection{Sample Selection}
The BCD classification covers a large range of
morphological types of ``host'' galaxies.  Within the
BCD class, there are a handful of galaxies which appear
to be hosted by dwarf ellipticals; that is, aside from
the central starburst, the underlying stellar distribution has
smooth, elliptical isophotes.  These BCD/dEs are
the {\it most likely} subclass of BCDs to look like dEs
after the starburst fades.  We thus selected a small sample
of BCD/dEs in order to investigate their gas kinematics and
distributions.  The galaxies in the present sample were selected 
from a large HI survey of blue compact dwarf galaxies \citep{SRWMB01}.

Physical properties of the selected galaxies are summarized in
Table \ref{tab:global}.  The distance to each system was calculated
from the systemic velocity using a Virgocentric infall model based on 
the outline of \citet{S80} and
an H$_0$ of 75 km s$^{-1}$ Mpc$^{-1}$.  Optical diameters
and luminosities were taken from the tabulation of \citet{SRWMB01}.
  The absolute blue magnitudes have been
corrected for Galactic extinction, but not for internal extinction or
for nebular contributions to the broadband luminosity.  All six
galaxies are at comparable distances and have similar luminosities.
Their oxygen abundances, however, span a wide range \citep{TMMMC91,IT99,S00}.
The L--band continuum flux listed in Table \ref{tab:global} was measured from the line--free channels
of the HI observations (see Section \ref{sec:hidata}); while the HI observations were
not optimized for continuum detection,  weak 20-cm continuum emission was
found associated with all six BCDs.

\subsection{HI Imaging Observations}
\label{sec:hidata}
Observations of UM 38, UM 323, MK 1418, MK 750, MK 900, and MK 324 were obtained with the 
VLA in its B and CS configurations during 1998--2000. As currently
implemented, the CS configuration is a modified version of the C 
configuration where a telescope from the middle of the north arm is 
re--positioned at an inner D configuration station to provide improved coverage
of the short spacings.  A summary of the observing sessions and the 
total on--source integration times are listed in Table \ref{tab:vlaobs}. 
During all observing sessions, the correlator was used in 2AD mode with 
the right and left circular polarizations tuned to the same frequency; 
the total bandwidth was 1.56 MHz.  The on--line
Hanning smoothing option was selected, producing final spectral data
cubes of 127 channels, each 2.6 \kms~wide.  Standard tasks in AIPS
\citep{AIPS}  were employed for calibration and 
preliminary data reduction.  Each set of observations was calibrated 
separately, using 3C 48 or 3C 286 as the flux and bandpass calibrator and nearby 
continuum sources as phase calibrators.  For all except MK 324,
continuum emission was removed with the AIPS task {\small \rm UVLIN} after the 
combination of {\it u--v}~data sets and prior to transformation to
the {\it x--y}~plane. 

The line data were transformed to the image plane with several weighting
schemes and combinations of configurations.  To check the data quality,
maps were made for each observing session.  The data cubes presented in this
paper were made from the combined data sets of the B and CS configuration 
observations.  A robust weighting technique was employed by the AIPS task 
{\small \rm IMAGR} to optimize the beam shape and noise levels \citep{B95}.  
The ``robustness parameter'' in {\small \rm IMAGR} controls the weighting of the {\it u--v}~data, 
permitting a fine--tuning between sensitivity and resolution in the final
data cubes.  As currently implemented, a robustness of 5 corresponds to 
natural weighting of the {\it u--v}~data (maximizing sensitivity at the cost of 
spatial resolution) while a robustness of --5 corresponds to
uniform weighting (maximal spatial resolution with lower sensitivity).
Additionally, some maps were made with {\it u--v}~tapers to increase their sensitivity
to low column density material.  The relevant {\small \rm IMAGR} parameters for a
selected sample of the maps are listed in Table \ref{tab:maps}.  Throughout
this paper we will refer to the lowest resolution, 
tapered data cubes as the ``tapered cubes'', to the robustness of
5 cubes as the ``natural weight cubes'', and to the robustness of
0.5 cubes as the ``intermediate weight cubes.'' 
All subsequent analysis was performed within
the GIPSY package \citep{GIPSY}.

The reduction stream for the MK 324 observations differed slightly from
the above synopsis.  First, line emission from MK 324 and
EXG 2323+1816 filled much of the bandpass, making it impossible to
remove the continuum emission in the {\it u--v}~plane.
A continuum image was created for each data cube by averaging the
few line--free channels; this image was then subtracted
from the original data cube.   Second, due to the low signal--to--noise of
the line emission in MK 324, the data were smoothed
during the transformation to the {\it x--y}~plane by averaging
2 consecutive channels, resulting in a cube of 63 channels, 
each 5.2 \kms~wide.  

To determine if the total flux density was recovered in the HI synthesis 
observations, total integrated profiles were constructed from the natural
weight data cubes after correcting for the primary beam shape with the
GIPSY task {\small \rm PBCORR} (Figure \ref{fig:flux}).  For all systems, 
the total flux density recovered in the VLA maps is in good agreement (within 10\%)
with previous single dish flux measurements \citep[e.g.,][]{HR86,vHG95,SRWMB01}.
The total HI masses listed in Table \ref{tab:global}
were calculated from the observed VLA flux density and the assumed distance to
the system.   

Moment maps of each data cube were computed in the following manner.
For each galaxy, both the tapered and natural weight cubes 
were smoothed to a resolution of twice the beam; the smoothed cubes were
clipped at the 2$\sigma$ level; the resultant clipped cubes were then 
interactively blanked to remove spurious noise spikes.
Signal was identified based on spatial continuity between channels.   A
conditional transfer was used to blank the corresponding locations in 
both of the original data cubes, and in the intermediate weight data cube,
after correcting for primary beam attenuation.
Moment maps of the tapered, natural, and intermediate weight data 
cubes were created with the GIPSY task {\small \rm MOMENTS} and are presented in 
Figures \ref{fig:um38}--\ref{fig:mk324}.

In the process of making moment maps, each data cube was searched for
possible companion galaxies.  Nearby systems were identified
in the lowest resolution data cubes of both UM 323 and MK 324
(Figures \ref{fig:um323comp} and \ref{fig:mk324comp}); we designate
these objects as EXG 0123-0040 and EXG 2323+1816, respectively.
Both EXG 0123-0040 and EXG 2323+1816 have good positional coincidences 
with faint optical galaxies.  Other HI synthesis observations of UM 323 
also identified an HI signal 
near the location of EXG 0123-0040 and at a similar velocity, designated 
as UM 323A  \citep{TBGS95}; while this original 
detection was subsequently 

\psfig{figure=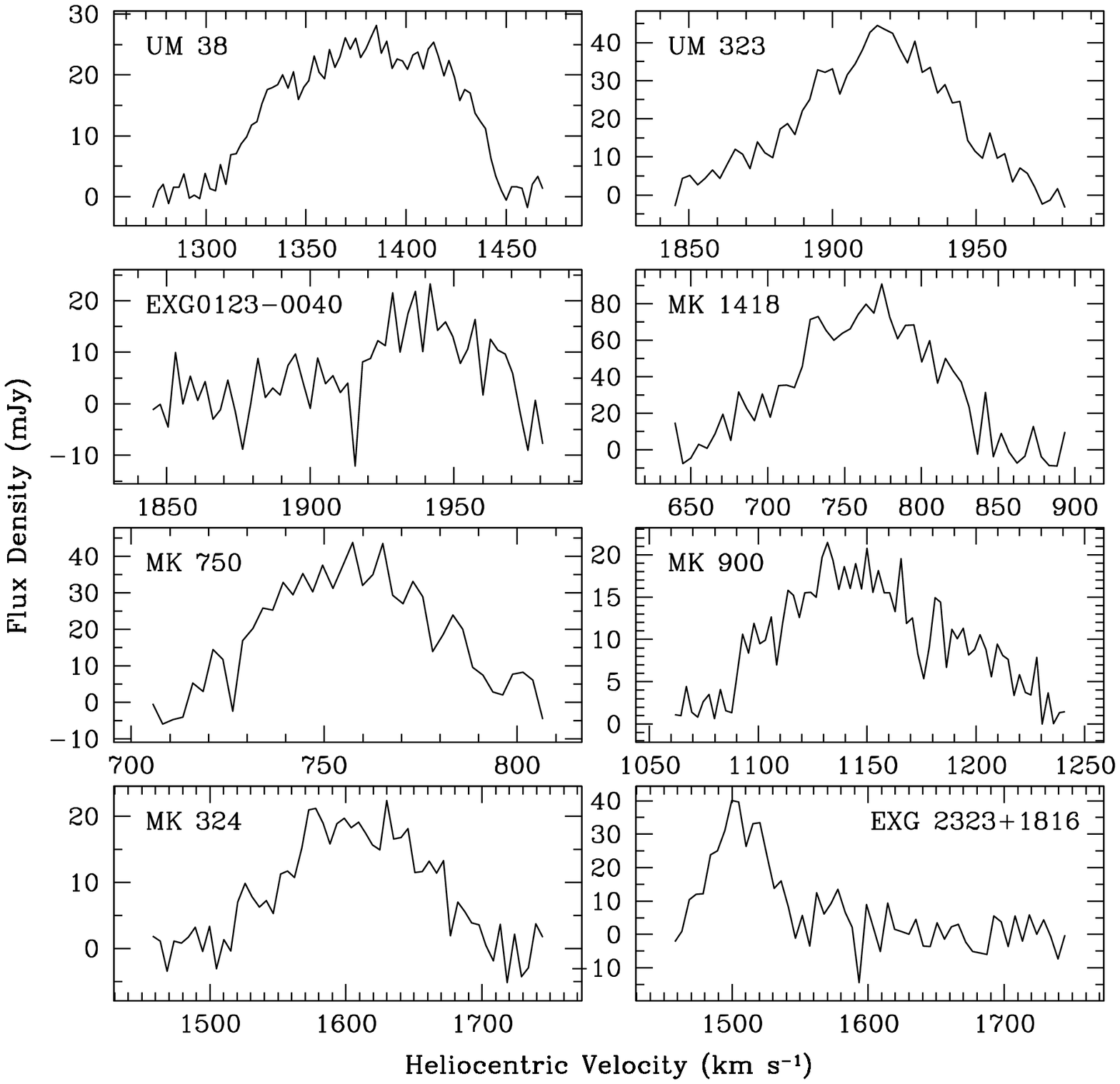,height=8.5cm,angle=0,bbllx=20 pt,bblly=160 pt,bburx=650 pt,bbury=710 pt,clip=t}
\figcaption[vanzee.fig01.ps] {Integrated HI profiles of the BCD sample
and of the two dwarf galaxies serendipitously discovered within the VLA primary beams. 
\vskip 10pt \label{fig:flux}}

\noindent
retracted  \citep[see][]{TBGS96},
this is probably the same object.  Note that the present data set
is completely independent from the previous D configuration observations of \citet{TBGS95}.
The integrated HI profiles of these 
systems are shown in Figure \ref{fig:flux};
the observed HI flux densities, systemic velocities, and velocity
widths are tabulated in Table \ref{tab:others}.  The primary beam 
correction may underestimate the total HI flux density of these systems since both of 
these galaxies are located slightly beyond the half power point of 
the main beam (see D$_{\rm center}$ in Table \ref{tab:others}). 
Both systems are at a projected distance of $\sim$130 kpc from the BCD, and have 
similar HI masses (30\% and 70\% of the target galaxy, respectively). 
These systems may not be either sufficiently massive, or close
enough, to have ``triggered'' the current burst of star formation in
UM 323 and MK 324, but the continuing detection of faint, but gas--rich,
systems which have been missed by optical surveys 
\citep[e.g.,][]{WLM96,WBDK97,PW99}
suggests that HI observations are a crucial step in
the process of determining the local environment of a galaxy.

\subsection{Optical Imaging}

Complementary optical images were obtained during several observing runs
at KPNO.\footnote{Kitt Peak National Observatory
is part of the National Optical Astronomy Observatories that are
operated by the Association of Universities for Research in Astronomy, Inc.
under contract to the National Science Foundation.} 
 H$\alpha$ and B--band images of UM 323, MK 750, MK 900, and MK 324 were obtained in 
1989--1990 with the KPNO No. 1 0.9m telescope; the observation and data reduction
procedures for these images were similar to those described in \citet{SAMGH91}.
H$\alpha$ and B--band images of UM 38, MK 900, and MK 324 were obtained with the
KPNO 0.9m telescope in June 1998; the observation and reduction procedures for 
these images are described in \citet{vZ00}.  Since no optical images were
available for MK 1418, or for the extended fields around UM 323 and
MK 324, images of these

\psfig{figure=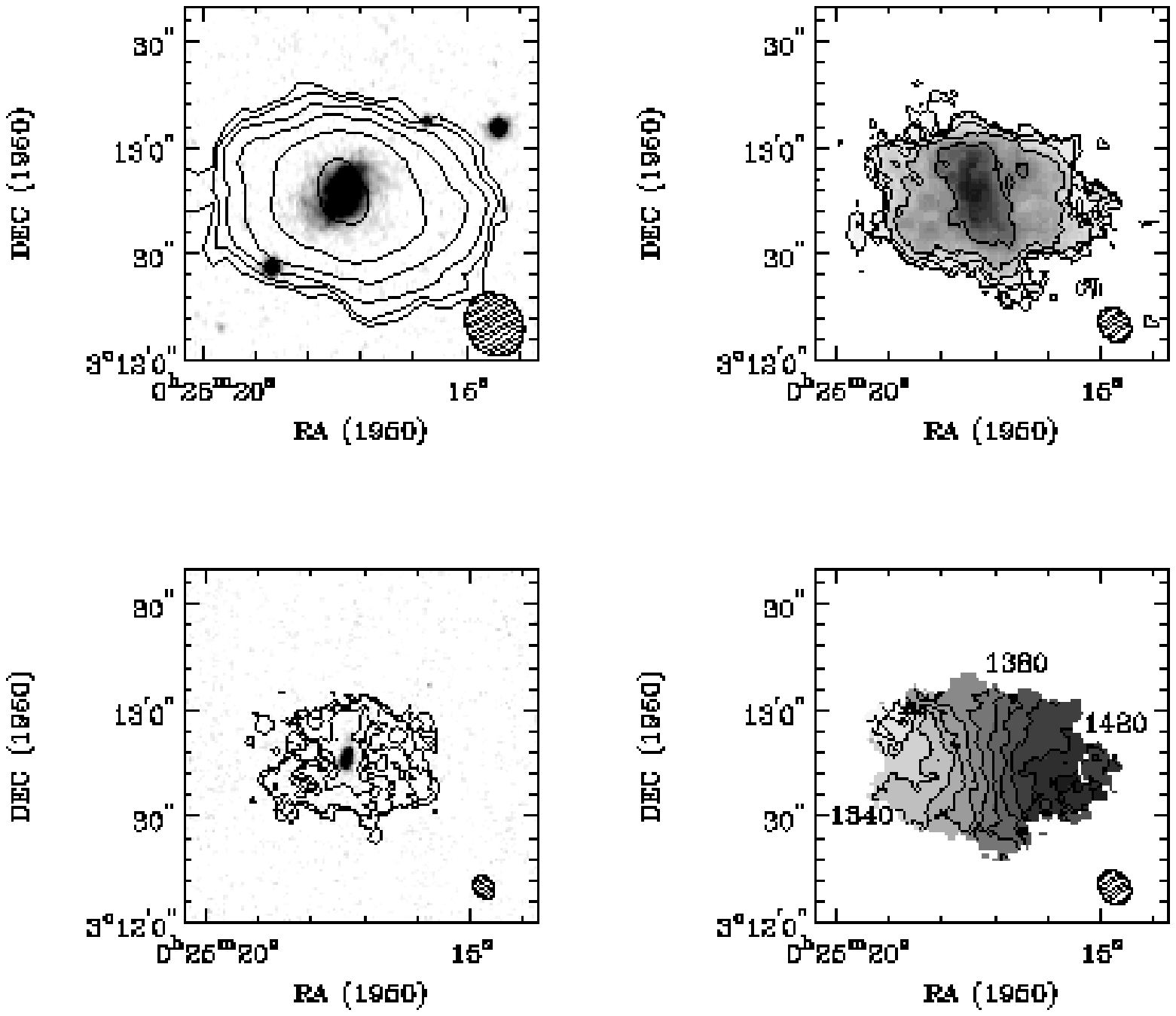,height=7.5cm}
\figcaption[vanzee.fig02.ps]{ Moment maps of UM 38.
(a) The HI contours from the tapered data cube are shown
overlayed on a B band image.  The HI contours
are 0.5, 1, 2, 4, 8, and 16  $\times$ 10$^{20}$ atoms cm$^{-2}$ with
a  beam size of  18.5 $\times$ 15.8 arcsec.  The pixel scale of the 
optical image is 0.688 arcsec pixel$^{-1}$.  
(b) The integrated intensity map from the natural weight data cube.
The HI contours are 1, 2, 4, 8, and 16 $\times$ 10$^{20}$ atoms cm$^{-2}$.
The beam size is 10.4 $\times$ 8.4 arcsec.
(c) The HI contours from the intermediate weight data cube overlayed on
an H$\alpha$ image.  The HI contours are 0.4, 0.8, and 1.6 $\times$ 10$^{21}$ atoms 
cm$^{-2}$ with a beam size of 7.0 $\times$ 5.4 arcsec.  The pixel scale 
of the H$\alpha$ image is 0.688 arcsec pixel$^{-1}$.
(d) The velocity field from the natural weight data cube.  
The contours are marked every 10 \kms.
\vskip 10pt
\label{fig:um38} }

\psfig{figure=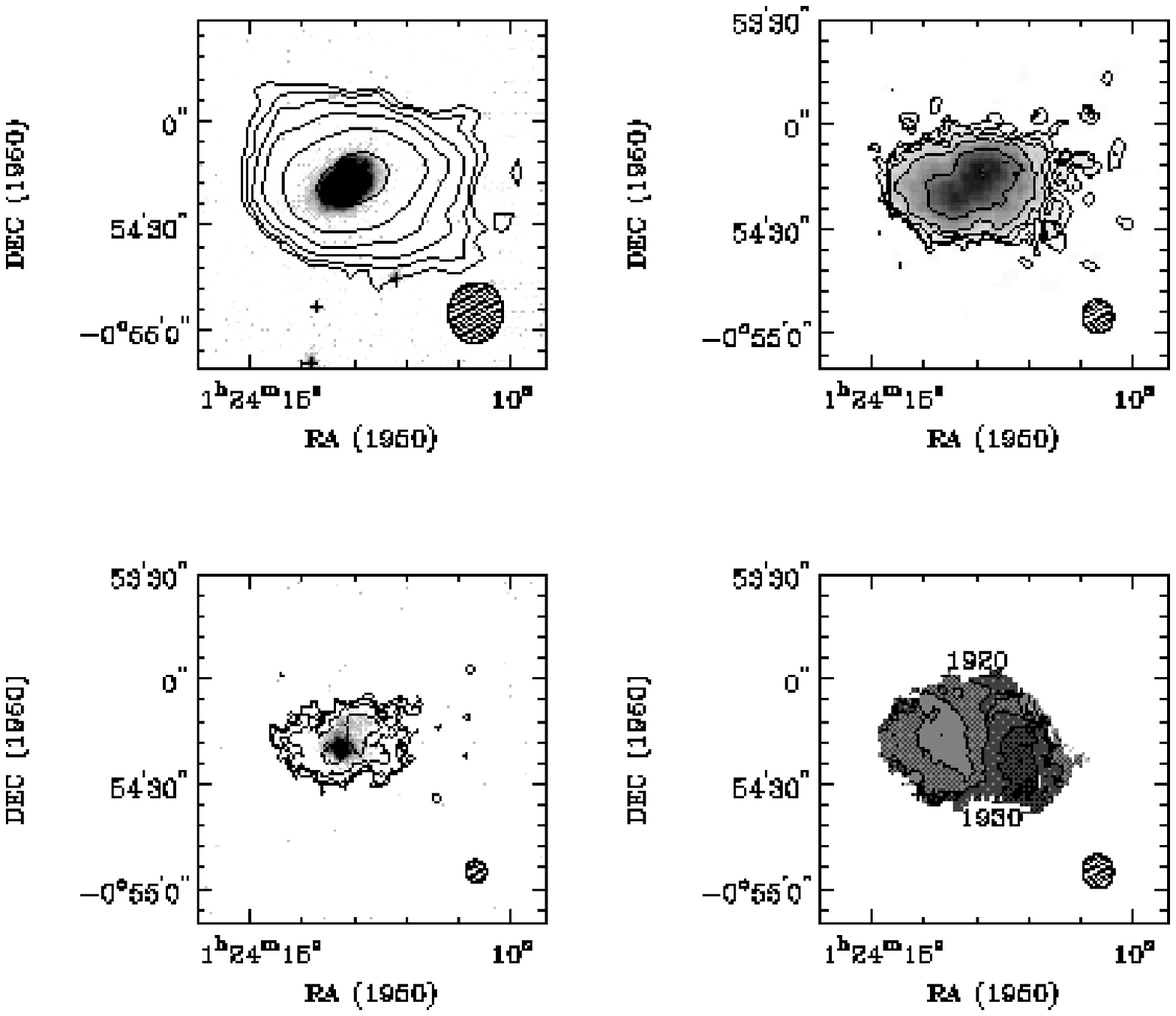,height=7.5cm}
\figcaption[vanzee.fig03.ps]{ Moment maps of UM 323.
(a) The HI contours from the tapered data cube are shown
overlayed on a B band image.
The HI contours are 0.5, 1, 2, 4, 8, and 16  $\times$ 10$^{20}$ atoms cm$^{-2}$
with a beam size of 17.4 $\times$ 15.6 arcsec.  The pixel scale of the 
optical image is 0.43 arcsec pixel$^{-1}$.  
(b) The integrated intensity map from the natural weight data cube.
The HI contours are 1, 2, 4, 8, and 16 $\times$ 10$^{20}$ atoms cm$^{-2}$.
The beam size is 9.6 $\times$ 8.8 arcsec.
(c)  The HI contours from the intermediate weight data cube overlayed on
an H$\alpha$ image.  The HI contours are 0.3, 0.6, 1.2, and 2.4 $\times$ 10$^{21}$ atoms cm$^{-2}$
with a beam size of 6.5 $\times$ 6.0 arcsec.  The pixel scale of the H$\alpha$ image
is 0.43 arcsec pixel$^{-1}$.
(d) The velocity field from the natural weight data cube.  The contours are marked 
every 10 \kms.
\vskip 10pt
\label{fig:um323} }

\psfig{figure=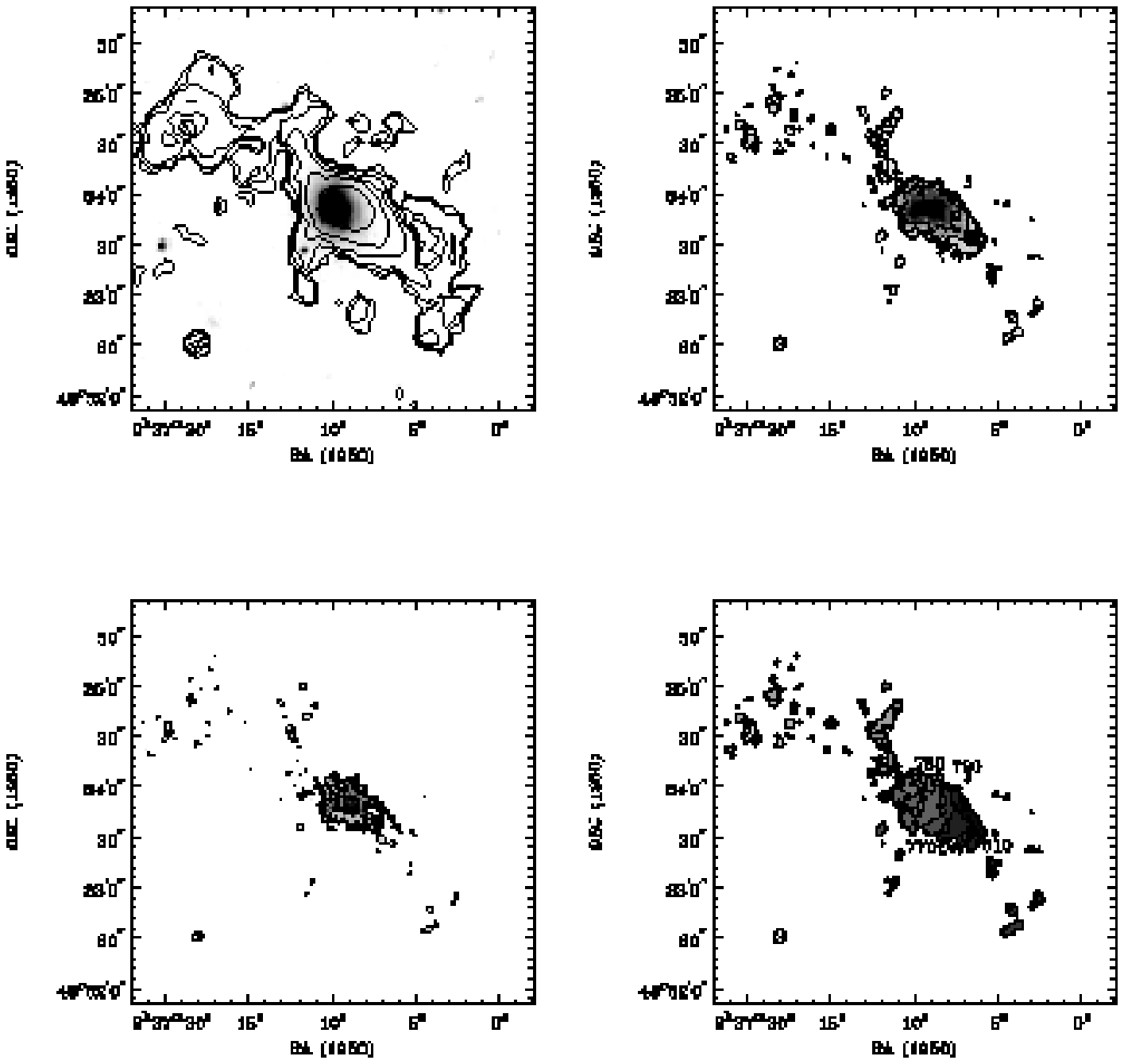,height=7.5cm}
\figcaption[vanzee.fig04.ps]{ Moment maps of MK 1418.
(a) The HI contours from the tapered data cube are shown
overlayed on an optical image from the Digitized Sky Survey.
The HI contours are 0.5, 1, 2, 4, 8, and 16  $\times$ 10$^{20}$ atoms cm$^{-2}$
with a beam size of 15.7 $\times$ 15.2 arcsec.  The pixel scale of the 
optical image is 1.008 arcsec pixel$^{-1}$.  
(b) The integrated intensity map from the natural weight data cube.
The HI contours are 1, 2, 4, 8, and 16 $\times$ 10$^{20}$ atoms cm$^{-2}$.
The beam size is 7.9 $\times$ 7.6 arcsec.
(c)  The integrated intensity map from the intermediate weight cube.  The
The HI contours are 0.3, 0.6, 1.2, and 2.4 $\times$ 10$^{21}$ atoms cm$^{-2}$
with a beam size of 5.6 $\times$ 5.5 arcsec. 
(d) The velocity field from the natural weight data cube.  The contours are marked 
every 10 \kms.
\vskip 10pt
\label{fig:mk1418} }

\psfig{figure=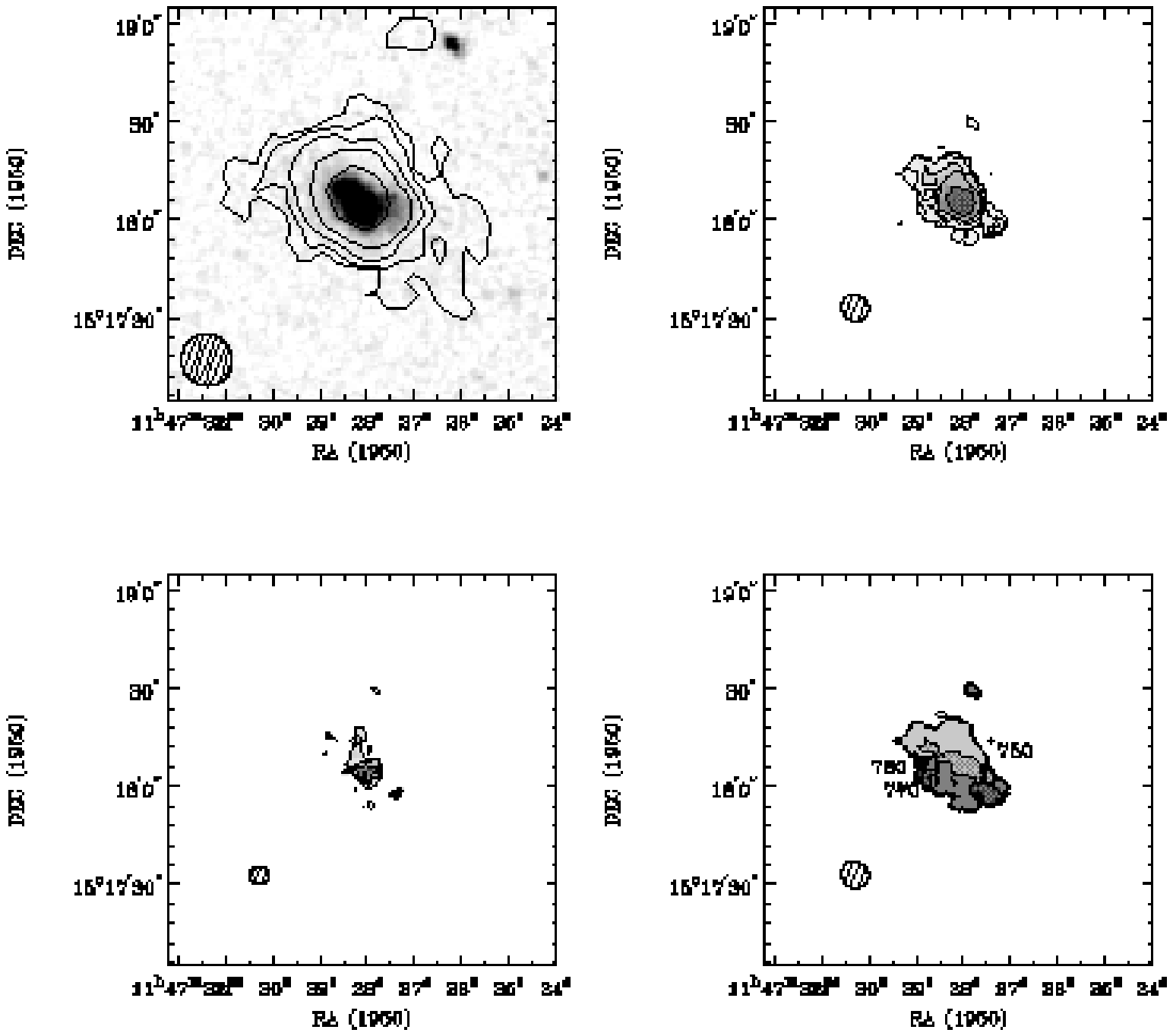,height=7.5cm}
\figcaption[vanzee.fig05.ps]{ Moment maps of MK 750.
(a) The HI contours from the tapered data cube are shown
overlayed on a B band image.
The HI contours are 0.5, 1, 2, 4, 8, and 16  $\times$ 10$^{20}$ atoms cm$^{-2}$
with a beam size of 16.1 $\times$ 15.5 arcsec.  The pixel scale of the 
optical image is 1.008 arcsec pixel$^{-1}$.  
(b) The integrated intensity map from the natural weight data cube.
The HI contours are 1, 2, 4, 8, and 16 $\times$ 10$^{20}$ atoms cm$^{-2}$.
The beam size is 8.7 $\times$ 7.7 arcsec.
(c)  The integrated intensity map from the intermediate weight cube overlayed on
an H$\alpha$ image. The HI contours are 0.3, 0.6, 1.2, and 2.4 $\times$ 10$^{21}$ atoms cm$^{-2}$
with a beam size of 6.0 $\times$ 5.4 arcsec.   The pixel scale of the H$\alpha$ image
is 0.43 arcsec pixel$^{-1}$.
(d) The velocity field from the natural weight data cube.  The contours are marked 
every 10 \kms.
\vskip 10pt
\label{fig:mk750} }

\psfig{figure=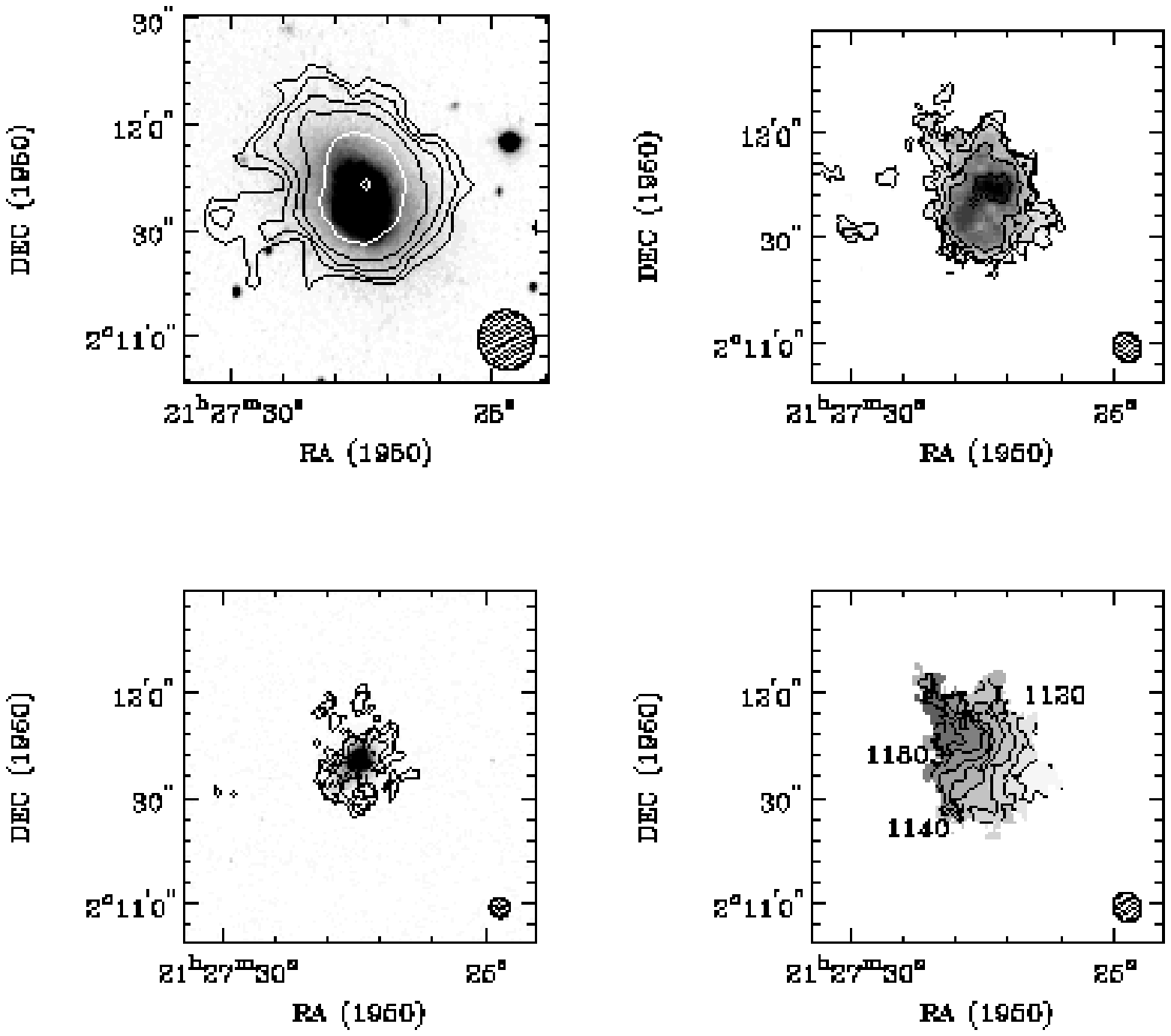,height=7.5cm}
\figcaption[vanzee.fig06.ps]{ Moment maps of MK 900.
(a) The HI contours from the tapered data cube are shown
overlayed on an B band image.
The HI contours are 0.5, 1, 2, 4, 8, and 16  $\times$ 10$^{20}$ atoms cm$^{-2}$
with a beam size of 17.0 $\times$ 15.9 arcsec.  The pixel scale of the 
optical image is 0.688 arcsec pixel$^{-1}$.  
(b) The integrated intensity map from the natural weight data cube.
The HI contours are 1, 2, 4, 8, and 16 $\times$ 10$^{20}$ atoms cm$^{-2}$.
The beam size is 8.6 $\times$ 7.8 arcsec.
(c)  The HI contours from the intermediate weight data cube overlayed on
an H$\alpha$ image.  The HI contours are 0.3, 0.6, 1.2, and 2.4 $\times$ 10$^{21}$ atoms cm$^{-2}$
with a beam size of 5.9 $\times$ 5.4 arcsec.  The pixel scale of the H$\alpha$ image
is 0.688 arcsec pixel$^{-1}$.
 (d) The velocity field of the natural weight data cube.  The contours are marked 
every 10 \kms.
\vskip 10pt
\label{fig:mk900} }

\psfig{figure=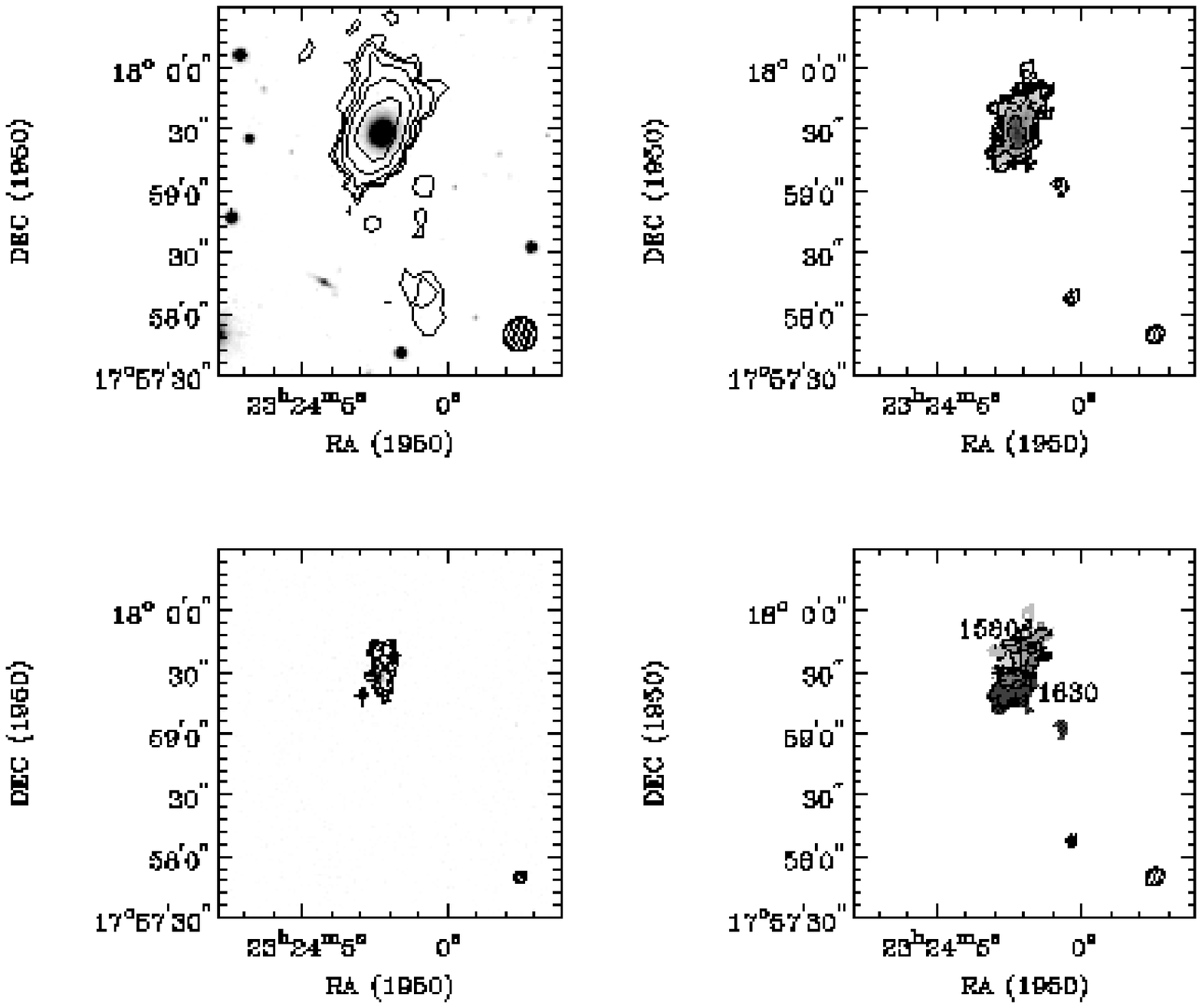,height=7.5cm}
\figcaption[vanzee.fig07.ps]{ Moment maps of MK 324.
(a) The HI contours from the tapered data cube are shown
overlayed on a B band image.
The HI contours are  0.5, 1, 2, 4, 8, and 16  $\times$ 10$^{20}$ atoms cm$^{-2}$
with a beam size of 16.7 $\times$ 15.7 arcsec.  The pixel scale of the 
optical image is 0.688 arcsec pixel$^{-1}$.  
(b) The integrated intensity map from the natural weight data cube.
The HI contours are 1, 2, 4, 8, and 16 $\times$ 10$^{20}$ atoms cm$^{-2}$.
The beam size is 9.4 $\times$ 7.7 arcsec.
(c) The HI contours from the intermediate weight data cube overlayed on
an H$\alpha$ image.  The HI contours are 0.4, 0.8, and 1.6 $\times$ 10$^{21}$ atoms cm$^{-2}$ 
with a beam size of 6.1 $\times$ 5.2 arcsec.  The pixel scale of the H$\alpha$ image
is 0.688 arcsec pixel$^{-1}$.
(d) The velocity field of the natural weight data cube.  The contours are marked 
every 10 \kms.
\vskip 10pt
\label{fig:mk324} }

\psfig{figure=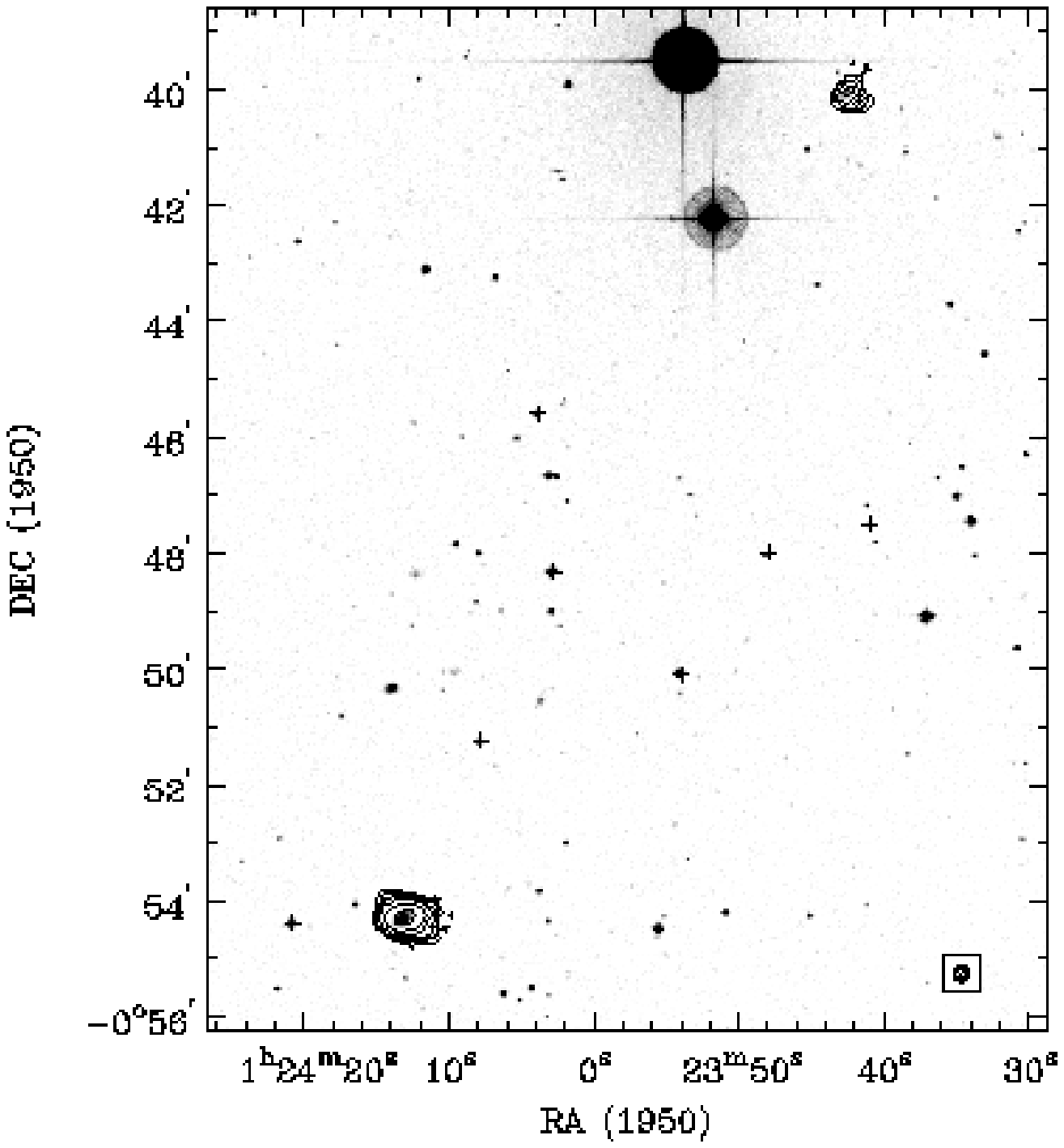,height=9.cm,angle=0,bbllx=100 pt,bblly=150 pt,bburx=650 pt,bbury=580 pt,clip=t}
\figcaption[vanzee.fig08.ps]{The HI contours from the tapered data cube are shown overlayed
on an image from the Digitized Sky Survey for UM 323 and EXG 0123-0040. 
HI image has been corrected for primary beam attenuation.
The HI contours are  0.5, 1, 2, 4, 8, and 16  $\times$ 10$^{20}$ atoms cm$^{-2}$
with a beam size of 17.4 $\times$ 15.6 arcsec. \label{fig:um323comp}}

\psfig{figure=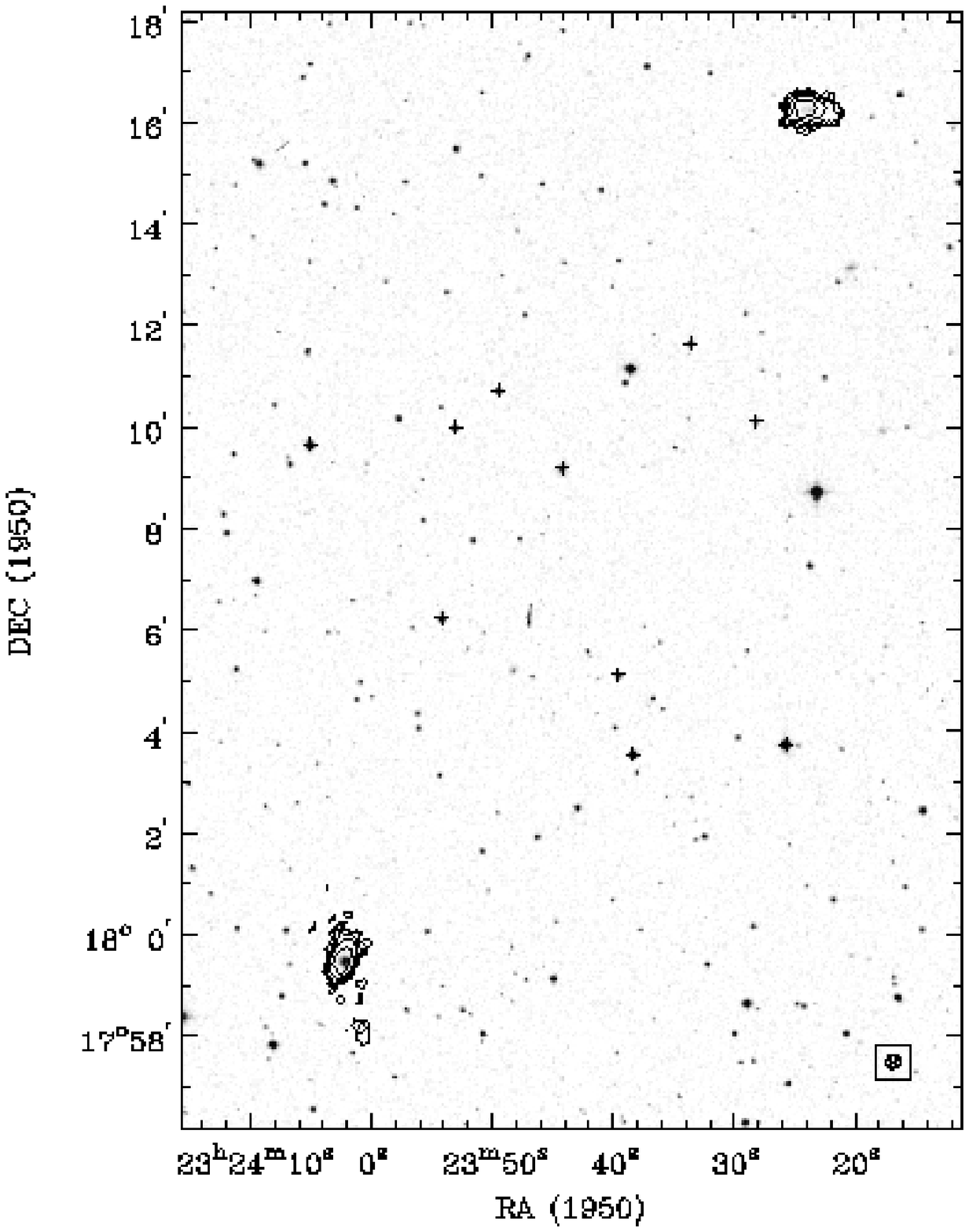,height=10.cm,angle=0,bbllx=100 pt,bblly=50 pt,bburx=650 pt,bbury=580 pt,clip=t}
\figcaption[vanzee.fig09.ps]{HI contours from the tapered data cube are shown overlayed
on an image from the Digitized Sky Survey for MK 324 and EXG 2323+1816.  The
HI image has been corrected for primary beam attenuation.
The HI contours are  0.5, 1, 2, 4, 8, and 16  $\times$ 10$^{20}$ atoms cm$^{-2}$
with a beam size of 16.7 $\times$ 15.7 arcsec. \label{fig:mk324comp}}

\newpage

\psfig{figure=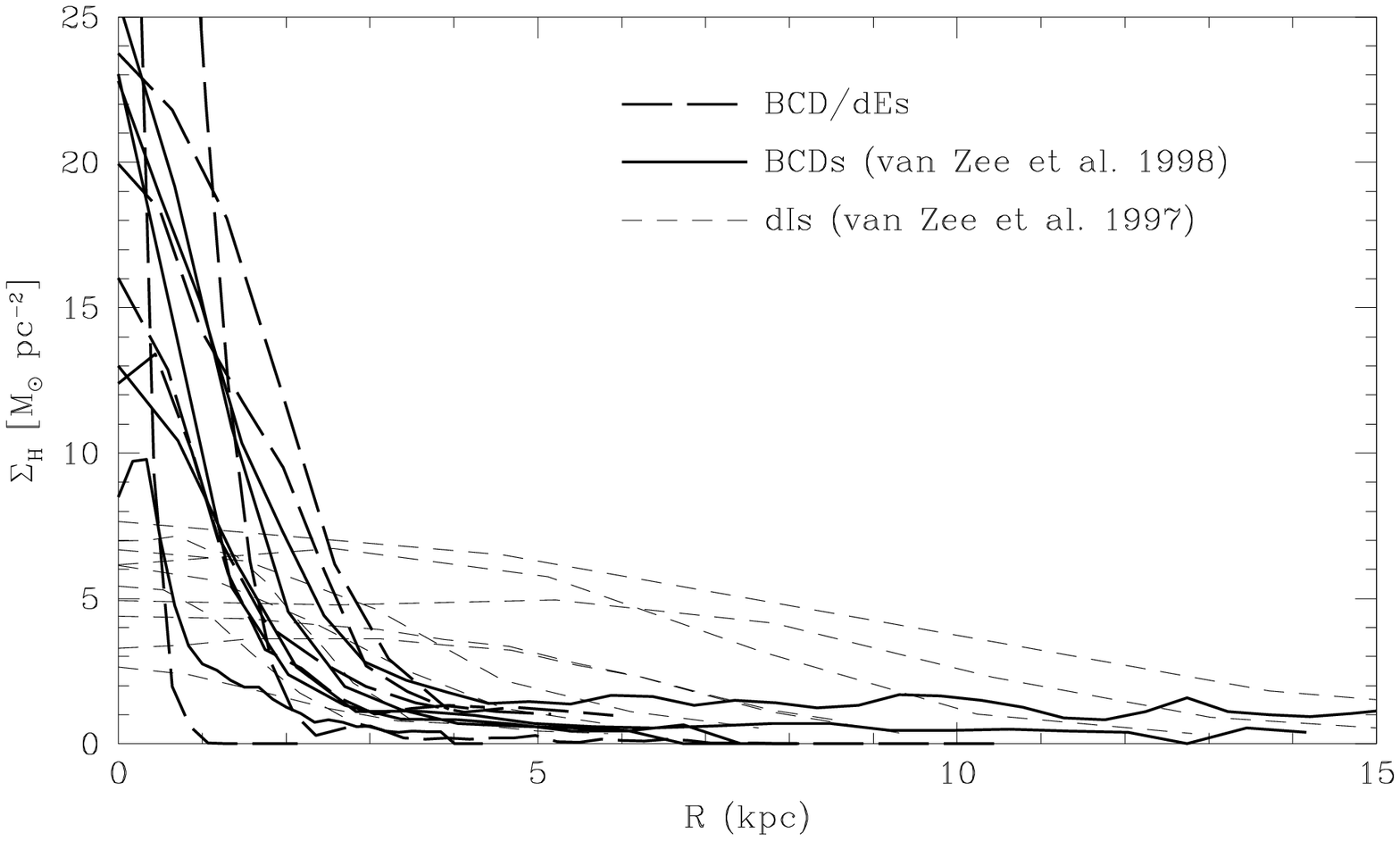,width=9.cm,angle=0,bbllx=0 pt,bblly=220 pt,bburx=600 pt,bbury=560 pt,clip=t}
\figcaption[vanzee.fig10.ps] {Azimuthal averages of the neutral gas distribution
of BCDs (van Zee et al.\ 1998a) and dIs (van Zee et al.\ 1997). The gas distribution
is more centrally concentrated in BCDs than in dIs.
\vskip 10pt
 \label{fig:gas}}

\noindent
 fields were extracted from the
Digitized Sky Survey\footnote{The Digitized Sky Surveys
were produced at the Space Telescope Science Institute under U.S. Government
grant NAG W--2166.}.  
Plate solutions for the optical images were derived from coordinates
of at least five stars listed in the APM catalog\footnote{http://www.ast.cam.ac.uk/$\sim$apmcat/} 
and are accurate to 0.5\arcsec.

Star formation rates were calculated from the H$\alpha$ images of UM 38, UM 323, MK 750,
MK 900, and MK 324.  First, the observed H$\alpha$ flux was corrected for internal 
and external absorption based on the observed spectroscopic Balmer line ratios \citep{TMMMC91,S00}.
 The derived H$\alpha$ luminosity was then converted to the current star--formation rate 
(SFR), using the conversion factor from \citet{K98}:
\begin{equation}
{\rm SFR} = 7.9 \times 10^{-42}~{\rm L(H\alpha)~M_{\odot}~yr^{-1}}.
\end{equation}
The current SFR and the gas depletion time scale (M$_{\rm HI}$/SFR) are tabulated in 
Table \ref{tab:global}.  In general, the current SFRs of the BCDs are quite high, with 
typical gas depletion times of approximately 1-5 Gyr.  However, it is important to
note that the star formation rate conversion factor used here may lead to an overestimate 
of the star formation rate in low metallicity galaxies, since the UV opacity will be 
lower in metal--poor stars.  Thus, it is possible that the gas depletion time scales
quoted here are underestimates.

\section{Results of HI Synthesis Imaging}

\subsection{HI Morphology}
\label{sec:hi}

The neutral gas distributions of the six BCDs are shown in Figures \ref{fig:um38}--\ref{fig:mk324}.
As has been seen in other observations of BCDs \citep[e.g.,][]{TBPS94,MCBF96,WBDK97,vWHS98,vSS98,PBMSK98},
the HI distribution extends well beyond the optical system.
The HI--to--optical diameters (measured at the 10$^{20}$ atoms cm$^{-2}$ and 25 
mag arcsec$^{-2}$ isophotes, respectively) are tabulated in Table \ref{tab:global}.
For all except MK 900, the neutral gas extends approximately 2 times the optical diameter,
which is quite typical for gas--rich galaxies \citep[see, e.g.,][]{BW94,BR97}.
Both MK 1418 and MK 324 have extended gaseous distributions which show kinematic peculiarities,
reminiscent of tidal tails.  Optically, these galaxies have smooth, regular isophotes, so it
is not clear if they are merger remnants, recently had an interaction, or if some other mechanism
produced their extended gaseous disks.  

\psfig{figure=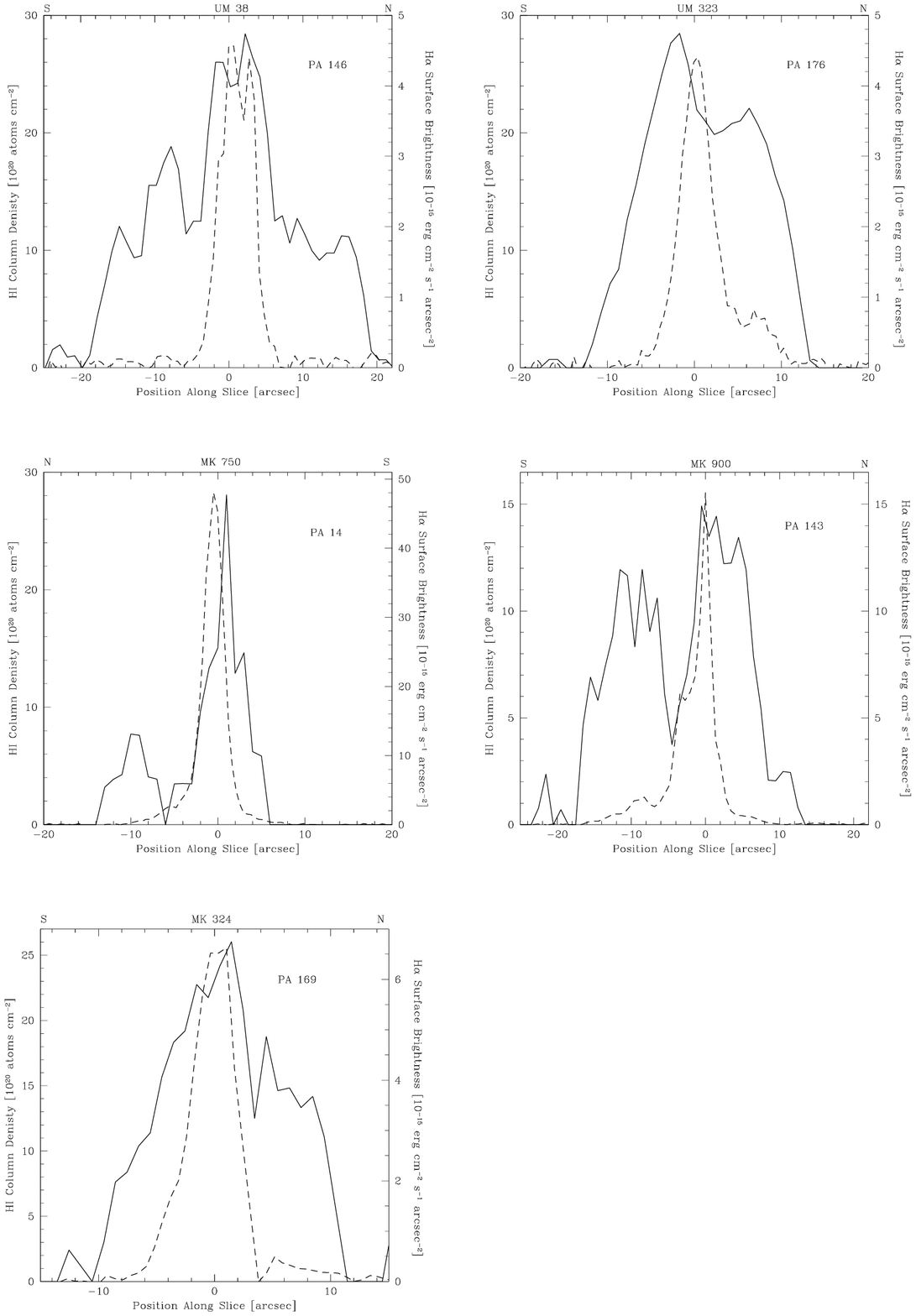,width=9.cm,angle=0,bbllx=0 pt,bblly=70 pt,bburx=600 pt,bbury=750 pt,clip=t}
\figcaption[vanzee.fig11.ps] {Comparison of the neutral (solid lines) and ionized gas (dashed lines) distributions
in BCDs.  In general, star forming regions are correlated with local peaks in the HI 
surface density.  However, in several cases, such as UM 38, the star forming region
appears to lie in a modest depression of the neutral gas, suggesting that
star formation has had an impact on the neutral medium. 
\vskip 10pt \label{fig:slice}}

Overall, the neutral gas distribution in these BCD/dEs is quite similar to that 
found in other BCDs.  Azimuthal averages of the gas distributions are shown in 
Figure \ref{fig:gas} for the present sample, a random sample of BCDs \citep{vSS98},
 and a sample of gas--rich dIs \citep{vHSB97}.  The gas is much more centrally concentrated in  the
BCDs [see also \citet{TBPS94} and \citet{SG00}], which presumably facilitates their high star formation rates.

More specifically, the neutral gas is concentrated in the regions of star formation
activity.  Slices through the neutral and ionized gas images of UM 38, UM 323, MK 750, 
MK 900, and MK 324 
are shown in Figure \ref{fig:slice}.  The position angle of each slice
was chosen so that the slice would intersect the brightest HII regions in each galaxy.
 As was seen with other BCDs \citep{vSS98},
the neutral gas density peaks coincide with regions of star formation.  However,
in the highest spatial resolution observations there
is some evidence of feedback between the star forming regions and the neutral
medium.  For instance, the brightest HII region in UM 38 is actually located
in a slight depression in the neutral gas surface density, as would be expected
if the neutral medium has been partially used, or ionized, by the star formation activity.

The observed peak column densities are tabulated in Table \ref{tab:colden}.
Since the observed column density depends on the beam dilution, the value
determined in each data cube is presented.  In general, the observed peak column densities 
are between 2--4 $\times$ 10$^{21}$ atoms cm$^{-2}$; when corrected
for inclination and neutral helium content, the peak neutral gas surface densities are
on the order of 3 $\times$ 10$^{21}$ atoms cm$^{-2}$ for all six galaxies. 
 Interestingly, these peak values are very similar to the predicted critical density as 
calculated from the \citet{T64} instability
criterion \citep[e.g.,][]{K89}.

\subsection{Neutral Gas Kinematics}
\label{sec:kin}

\subsubsection{Velocity Fields}
\label{sec:vel}

The velocity fields are shown in Figures \ref{fig:um38}d--\ref{fig:mk324}d.
In all cases, a clear 
velocity gradient is present; interestingly, the
velocity gradient is aligned along the optical major axis of only MK 900 and
MK 324.   Both UM 38 and MK 900 have relatively smooth, ordered velocity fields while those
for UM 323, MK 1418, MK 750 and MK 324 are significantly more complicated.   Less than
perfect ordered rotation has also been found in other high spatial resolution
observations of BCDs \citep[e.g.,][]{vWHS98,vSS98}.
Nonetheless, despite the kinematic complexities, it is clear that these systems all have 
large rotation velocities.

\subsubsection{Rotation Curves}

Model rotation curves for the BCDs were derived from  tilted--ring
analysis of the observed velocity fields using the GIPSY task {\small \rm ROTCUR}.
The process of fitting a rotation curve is iterative.  First, the
kinematic center and systemic velocity were determined using both sides
of the velocity field to constrain these global parameters.  Next,
the center and systemic velocity were held fixed while the
inclination angle and position angle were derived.
Unfortunately, the complex nature of the velocity fields imply a large
degree of uncertainty in the kinematic parameters; for all except UM 38, the derived 
kinematic parameters (Table \ref{tab:hiparms}) should only be considered rough estimates. 
In particular, it was not possible to constrain the inclination angle based on 
the velocity field alone; an estimate of the inclination angle was derived 
from the observed axial ratios (both HI and optical).  Finally, despite the 
apparent solid body nature of the velocity fields, the final rotation
curves produced by {\small \rm ROTCUR} had a tendency to flatten
in the outer parts; thus, in the end, only UM 38 was satisfactorily fit
by this method.

Nonetheless, the {\small \rm ROTCUR} fits were sufficient to indicate the
major axis of each galaxy so that position--velocity diagrams could be
created (Figure \ref{fig:pv}).  As expected from the velocity fields,
the p--v diagrams indicate solid body rotation throughout the gas disk
for UM 323, MK 1418, MK 750, MK 900, and MK 324, while UM 38 appears to be differentially
rotating in the outer disk.  Also shown in Figure \ref{fig:pv}
is the expected observed rotation curve from the {\small \rm ROTCUR} model
for UM 38, and estimates of solid body rotation curves for the other
galaxies.  In fact, the p--v diagrams  provide  
a good estimate of the rotation curves for UM 323, MK 1418, MK 750, 
MK 900, and MK 324, assuming that no warps or other kinematic abnormalities are present. 
This is probably a reasonable assumption for MK 900, but the derived
kinematic parameters for the other four galaxies are significantly more uncertain.

The radii and velocities (corrected for inclination) of the last measured 
points of the rotation curves are tabulated  

\psfig{figure=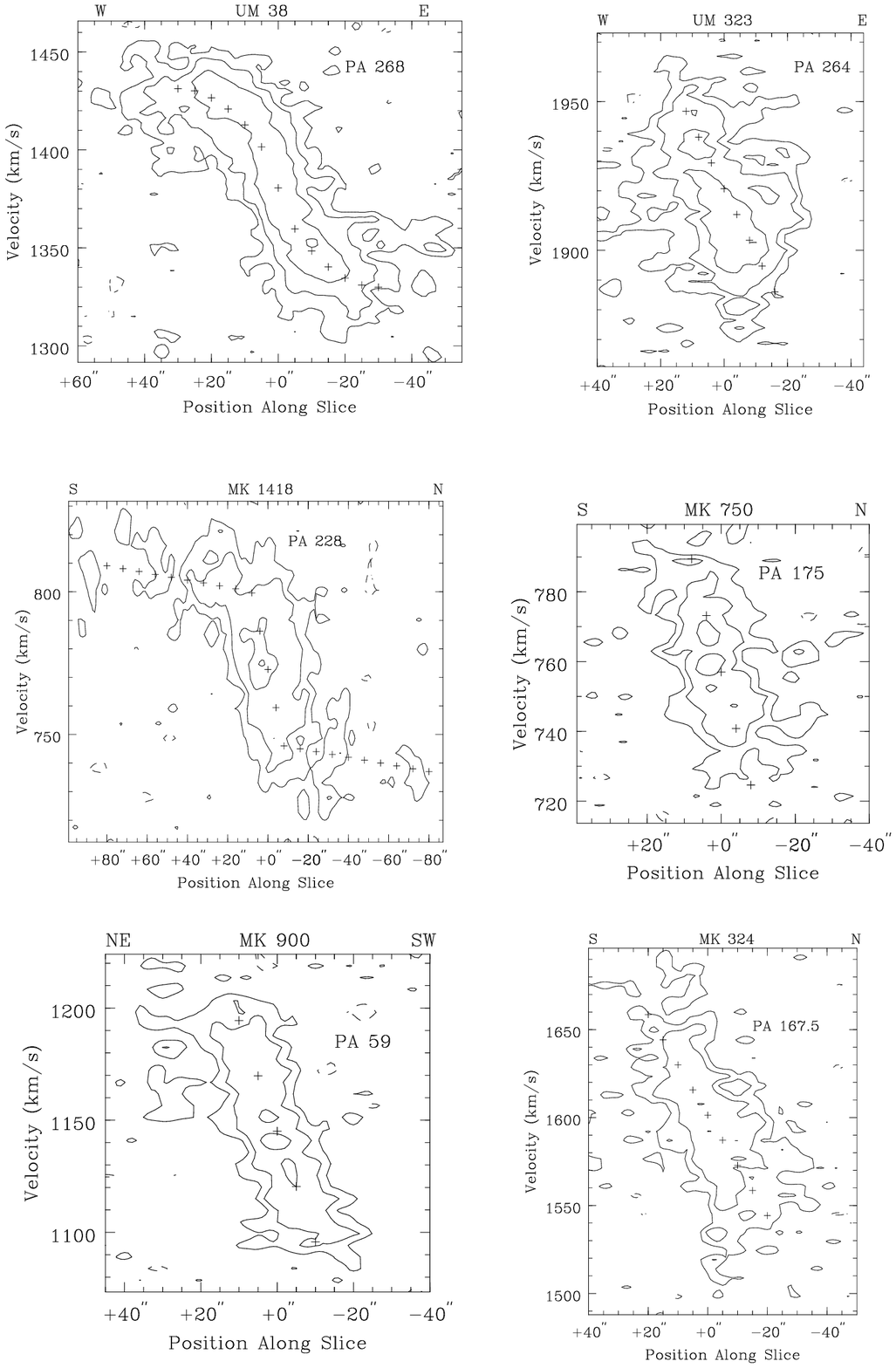,width=9.cm,angle=0,bbllx=0 pt,bblly=100 pt,bburx=600 pt,bbury=750 pt,clip=t}
\figcaption[vanzee.fig12.ps] {Position--Velocity diagrams from the natural weight data cubes; 
the UM 38 and MK 900 data cubes have been hanning smoothed.  The contours represent 
-3$\sigma$, 3$\sigma$, 6$\sigma$, and 12$\sigma$.  A rotation curve as derived 
from tilted--ring models is superposed on the p--v diagram for UM 38; in the other
five panels, crosses denote the adopted solid body rotation curves. 
\vskip 10pt
 \label{fig:pv}}

\noindent
in Table \ref{tab:hiparms}.
Since the rotation curves were poorly constrained for all except UM 38 and MK 900,
these values are only rough estimates.  If the BCDs
are dark matter dominated, as has been found for virtually all dwarf
galaxies studied \citep[e.g.,][]{CB89}, 
the dynamical mass can be calculated from the derived rotation curves.
For a spherical mass distribution (appropriate for most dark matter halos), 
the dynamical mass within radius $r$ is:
\begin{equation}
M_T(r) = 2.326 \times 10^5~V^2(r)~r
\end{equation}
where $V(r)$ is in km s$^{-1}$ and $r$ is in kpc.  The derived dynamical
masses for the last measured point of the rotation curves 
are tabulated in Table \ref{tab:hiparms}.  The HI--to--dynamical
mass ratios for UM 38, MK 1418, MK 750 and MK 900 are quite typical for dwarf galaxies
in general \citep[e.g.,][]{SBMW87,vHSB97}
while those of UM 323 and MK 324 are significantly higher than average.
However, both UM 323 and MK 324 had extremely complex velocity fields, so
this may simply reflect on the poorly derived rotation curves, rather than
an intrinsic difference in galaxy properties.

\subsubsection{Thermal Motions}
\label{sec:disp}

An estimate of the neutral gas velocity dispersion was obtained from 
Gaussian fits to the spectral profiles in the natural weight data cubes.
The median $\sigma$ is tabulated in Table \ref{tab:hiparms}.
The observed velocity dispersions are quite typical of dwarf galaxies
\citep[e.g.,][]{YL96,vHSB97}.
In all cases, the observed velocity dispersions are significantly less than
the rotational component in these galaxies, indicating that these systems
are rotation dominated.

\section{Discussion}
\label{sec:dwfs}

\subsection{The Dynamics of BCDs}

Based on these and other observations of BCDs and dIs \citep[e.g.,][]{vHSB97,vSS98,S99},
it appears that all types of moderate luminosity gas--rich dwarf galaxies are 
rotation dominated systems.  In particular, as illustrated in Figure \ref{fig:pv},
the BCD/dEs show significant velocity gradients across the optical and HI disks, which   
 implies that the gaseous disks have significant angular momenta.
The specific angular momenta for a given radius 
in the disk
can be calculated from the derived rotation curves:
\begin{equation}
J/M = V \times R
\end{equation}
where $J/M$ is the specific angular momenta, $V$ is the rotation velocity, 
and $R$ is the radius.  It is difficult to specify a ``characteristic'' value for 
the specific angular momentum of the disk in low mass galaxies because the specific
angular momentum increases with radius for galaxies undergoing solid--body rotation.
 However, it is useful to compare
the specific angular momenta of the outermost neutral hydrogen gas in BCDs and
low surface brightness (LSB) dIs.  The specific angular momenta of the extended
HI disks were computed from the maximum rotation velocities and radii tabulated
in Table \ref{tab:hiparms} (BCDs) and in Table 4 of \citet{vHSB97} (LSBDGs). 
 While the definition of ``the outermost'' gas
is sensitivity dependent, the two studies shown in Figure \ref{fig:angmom}
have similar sensitivities, and the galaxies have similar HI--to--optical ratios.
As illustrated in Figure \ref{fig:angmom}a,  the BCDs have lower specific angular momenta 
than the LSB dIs, as one would expect for rotation curves which reach
similar maximum velocities at smaller radii (Figure \ref{fig:angmom}b).
Since the rotation curves of dwarf galaxies are primarily solid body, the fact
that BCDs reach a similar maximum velocity within a smaller radius implies
that BCDs have steeply rising rotation curves.

The fact that BCDs appear to have steeply rising rotation curves (low angular momenta) 
has several interesting consequences for evolutionary studies of BCDs.  First, the shape 
of the rotation curve suggests that the matter distribution is centrally
concentrated \citep[see also][]{MSSK98}.  As expected for galaxies which are 
classified as ``compact,'' studies of their stellar distributions indicate that 
the stars are centrally concentrated in BCDs \citep{SN99,NS00}; similar results are found
for the gaseous distribution as well (Figure \ref{fig:gas}).
However, the rotation curve results also suggest that the dark matter distributions
may be centrally concentrated in BCDs.  Given the poor constraints on the
derived rotation curves in this sample, we have elected to forgo extensive
analysis at this time (i.e., dark matter decompositions).  However, more extensive
analysis could place strong constraints on dark matter models, since
dwarf galaxies are usually dark matter dominated.  BCDs clearly inhabit an
important, and as yet unexplored, region of parameter space.

A second major implication from the observed rotation curve shapes (angular momenta) 
is that BCDs have higher threshold densities for the onset of star formation.
Using the classic \citet{T64} instability analysis for a rotating 

\psfig{figure=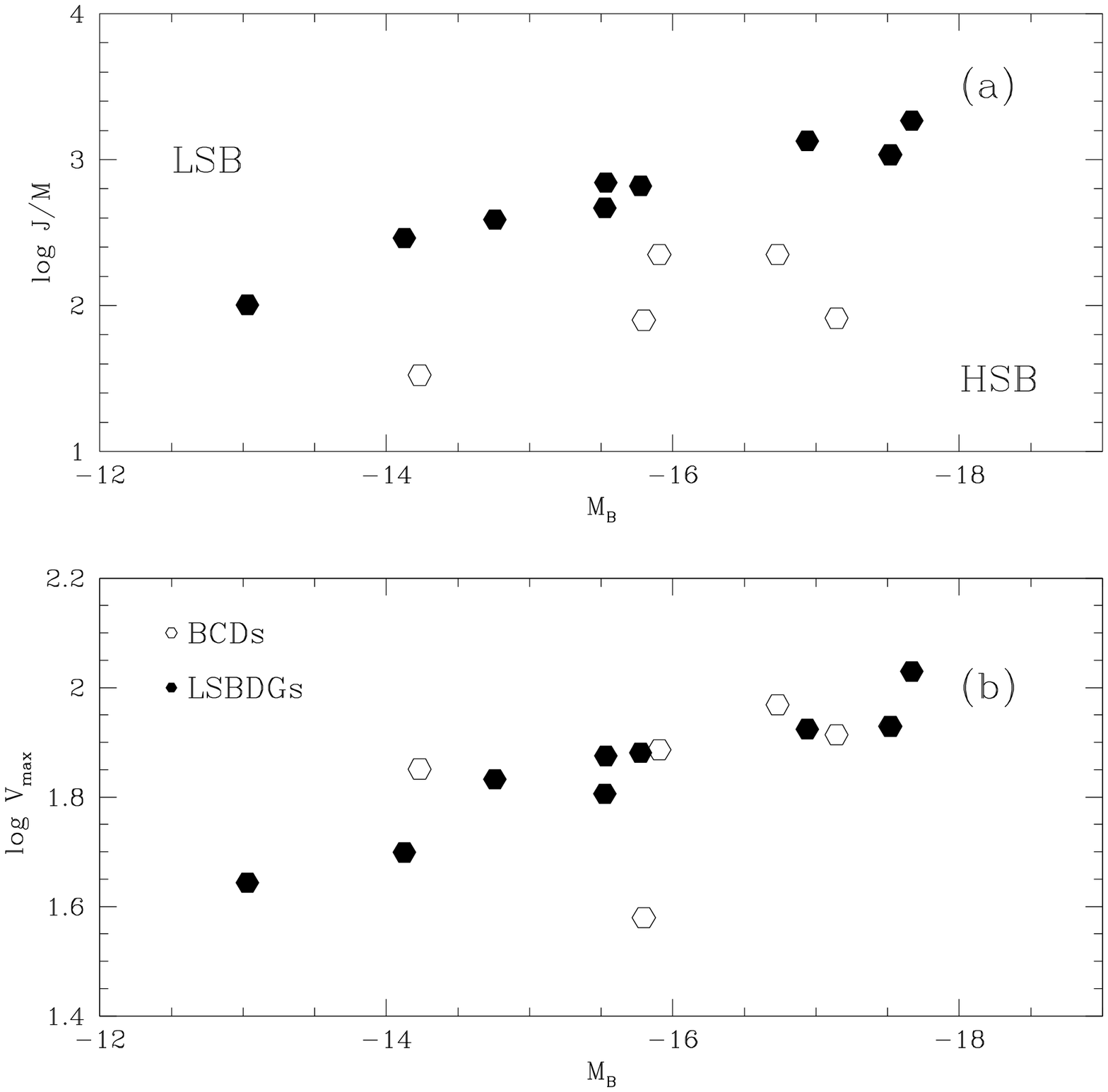,width=9.cm,angle=0,bbllx=0 pt,bblly=170 pt,bburx=600 pt,bbury=700 pt,clip=t}
\figcaption[vanzee.fig13.ps]{ (a) Specific angular momenta of the outer gas disks in BCDs (open
circles) and  LSB dIs (filled circles, van Zee et al.\ 1997).  The BCDs have lower specific
 angular momenta compared to comparable luminosity LSB dIs. (b) Maximum rotation velocities
of the BCD and dI samples.  Both types of galaxies have similar rotation velocities.
The fact that both BCDs and LSB dIs follow the same luminosity--linewidth relation has 
implications for galaxy evolution models:  BCDs have steeper rotation curves than the LSB dwarfs,
and thus have lower angular momenta and higher critical densities for star formation.
\vskip 10pt
\label{fig:angmom}}

\noindent
disk, the
threshold density for star formation can be derived from the balance
between gravitational collapse and  the effects of both thermal pressure and rotational shear.
For a solid body rotation curve, this type of analysis results in a star formation
threshold density which is proportional to the slope of the rotation curve.
In other words, a galaxy with a steep rotation curve (low specific angular momenta) 
will have a higher threshold density for star formation.   While the
instability threshold is a global dynamical property, 
it is  particularly important to note that the current observed
gas column densities appear to hover around the instability threshold densities
in both the BCDs \citep[this paper and][]{vSS98}
and dIs \citep{vHSB97}, and that the local column density peaks are correlated
with sites of current star formation activity.  Thus, there appears to be a 
strong connection between crossing the global instability threshold and the onset 
of star formation, a local process.

Angular momentum appears to play a crucial role in determining the
ultimate star formation history of a low mass galaxy.  The gas associated with
a galaxy with high angular momentum will have difficulty collapsing, and thus at the 
present day it will appear as a diffuse, low surface brightness galaxy.  On
the other hand, the gas associated with galaxies with low angular momenta will 
be able to collapse further, leading to smaller, compact, high surface brightness
galaxies.  Concurrently, in these compact systems, the onset of star formation may be delayed
because the threshold density is extremely high.  For example, if the infalling
gas accretes at a steady rate, it will take longer for the gas density to exceed
the critical density in a galaxy with lower angular momenta.  Further, once the
gas density does exceed the critical density, there will be a large fuel supply 
readily available for star formation.  It is thus quite likely that these low angular
momenta, high  density systems will be more susceptible to a ``burst'' mode
of star formation than  similar mass, but high angular momenta, low density systems.
It is also logical to hypothesize that starbursts will occur episodically in these
compact systems, with the starburst duty cycle regulated by the frequency at which 
the central gas density reaches the critical density for star formation.

\subsection{Dwarf Galaxy Evolutionary Scenarios}

As mentioned in the introduction, one of the key questions in dwarf galaxy
evolution is the ultimate fate of the star--bursting dwarf galaxies.
Since the starburst phase in BCDs is expected to be a transitory event which 
lasts only a few 100 Myr, one logical hypothesis is that BCDs must evolve 
from (and into) other types of low mass galaxies.  Several of the major unresolved 
issues in dwarf galaxy evolutionary scenarios are: (1) do all dwarf galaxies undergo 
a starburst phase? (2) what mechanisms initiate or trigger a starburst event? 
(3) what mechanisms terminate a starburst event? and (4) how does the galaxy
evolve after the burst phase?
 
For the last 30 years, most discussions of dwarf galaxy evolutionary scenarios
have centered around the morphological similarities between starbursting 
dwarf galaxies, dwarf irregular galaxies, and dwarf elliptical galaxies
\citep[e.g.,][]{LF83,K85,LT86,DH91,J94,PLFT96,SHRCK98}.
These studies have shown that the stellar distributions of all three types of 
low mass galaxies are well fit by exponential profiles, with a few notable exceptions
\citep[e.g.,][]{DCPST97,DCC99}.  Thus, based on the stellar distributions alone,
it has been argued that BCDs can evolve into either the
dwarf elliptical class or dwarf irregular class with only minor changes in
the stellar distribution and optical morphology.  Arguments against such passive morphological
evolution usually focus on the fact that the stellar distributions of BCDs
are much more compact than the typical dwarf irregular or dwarf elliptical galaxy;
however, since there is some overlap in the distribution functions for all three type of galaxies,
such morphological evolution cannot be excluded based on the stellar distributions
alone.   

In the next two sections we outline the arguments for and against the canonical 
BCD evolutionary scenarios.  First, in Section 4.2.1, we discuss whether
 gas--rich BCDs can evolve into gas--poor dwarf ellipticals.  Second, 
in Section 4.2.2, we discuss whether post--burst BCDs can be found within the
general population of gas--rich dwarf irregular galaxies.  A similar discussion
with a focus on the evolution of dwarf irregular galaxies may be found in \citet{SB95}. 

\subsubsection{BCD to dE: Quenching the Burst}

Secular evolution from BCD to dE requires more than just fading of the central
starburst.  All of the dwarf ellipticals in the Local Group are gas--poor
\citep[with M$_{\rm HI} < 10^{5} M_{\odot}$,][]{YL97,Y99,Y00}, while the BCDs are gas--rich
systems, with M$_{\rm HI} > 10^{7} M_{\odot}$.  Thus, evolution from BCD to dE
requires either efficient gas consumption or removal of the remaining interstellar medium.
The latter process can be accomplished in at least two distinct ways: 
(1) \citet{DS86} proposed that a starburst might have sufficient kinetic energy to 
disrupt the interstellar medium in a low mass galaxy, and 
(2) \citet{LF83} proposed that the ISM might be removed via
ram pressure stripping. 

Hydrodynamic models initially indicated that a starburst could have a disastrous
effect on the interstellar medium within a low mass galaxy \citep[e.g.,][]{DS86,DG90}; 
however, recent 
models, which include a large dark matter component, have shown that it is much more 
difficult to remove the interstellar medium with a single starburst than previously 
thought \citep[e.g.,][]{DH94,ST98,MF99}.  Rather, while some of the hot, ionized gas may 
escape \citep[as possibly seen in NGC 1569,][]{HDLFGW95}, the neutral medium is largely unaffected
by the starburst.  Further, detailed star formation histories of the Local Group dEs
indicate that these dwarf galaxies appear to have had
several discrete star--formation episodes \citep{M98}, rather than the single burst
envisioned by \citet{DS86}.   Thus, it now appears 
that it is unlikely  that a single starburst will completely use up, or blow--out, the 
interstellar medium of a dwarf galaxy. 

Ram pressure stripping, on the other hand, is still a viable mechanism to remove 
the remnants of the ISM in a post--burst low mass galaxy.  Ram pressure stripping
will occur predominantly in high density regions, the exact type of environment
favored by dwarf elliptical galaxies.  However, this process will operate regardless
of the star formation activity in the low mass galaxy; thus, while this is a viable
mechanism to transform a gas--rich star forming galaxy into a gas--poor evolved
galaxy, it is tangential at best to the main discussion of the evolutionary fate of
star--bursting dwarf galaxies.

The neutral gas kinematics presented here and in \citet{vSS98} provide the first
definitive evidence that the majority of BCDs will not evolve passively into dwarf
elliptical galaxies.  All of the BCDs studied so far are rotation dominated and
have significant angular momenta.  In contrast, dwarf elliptical galaxies are
not rotationally supported \citep{BN90,BPN91}.   
Of the 5 dwarf ellipticals observed, only one has a maximum stellar
rotation velocity $>$ 15 \kms; the remainder are consistent with zero stellar rotation.
As mentioned previously, the BCDs in the present
study were selected specifically to have optical properties similar to dwarf elliptical
galaxies, since these would be the {\em most likely} to be classified as dwarf
ellipticals after the starburst; however, even these galaxies have significant
rotation velocities.  Thus, these observations indicate that BCDs cannot evolve
passively into dwarf elliptical galaxies since such a morphological evolution 
would require the loss of angular momentum.

Further, the possibility that ram pressure stripping may be a necessary process to
remove the interstellar medium during the formation of dwarf elliptical galaxies leads
directly to another strong argument against  generic morphological evolution
from a starbursting dwarf galaxy to a dwarf elliptical \citep[such as
the one envisioned by][]{DS86}: BCDs and dEs have very different clustering properties.
Dwarf ellipticals are generally found in clusters or as companions to massive
galaxies while BCDs are typically field objects.  In fact, dEs are among  the
most strongly clustered galaxies known \citep{FS89}, while BCDs are among the least clustered
\citep{IMS88,RSM94,PULTG95,LSLR00}.  Thus, it may be necessary to consider the evolution of
field and cluster dwarf galaxies separately:  BCDs formed in the field will probably retain
a substantial fraction of their ISM, even after the starburst phase, and thus
will be classified as dwarf irregular galaxies if identified either pre-- or post--burst; 
meanwhile, some BCDs formed in cluster environments may evolve into dwarf elliptical 
galaxies, but such morphological evolution will not occur via passive secular evolution.

\subsubsection{BCD to dI: Quelling the Burst}

One of the difficulties for discussions of BCD to dI evolutionary scenarios is
a nomenclature issue: in many regards, BCDs {\em are} dIs.   Since the dwarf 
irregular classification is usually based on irregular morphology and gas--richness, 
star--bursting dwarf galaxies are included in this more general
classification.  Thus, rather than true morphological evolution, the BCD to dI 
evolutionary scenario is a matter of degree: can post--burst BCDs be found within 
the general dwarf irregular class, and also, is a starburst phase common 
to all gas--rich low mass galaxies? 

These questions have been addressed extensively by \citet{vZ00,vZ01}, who argues
that the majority of dIs do not go through a starburst phase.  Rather, the 
star formation histories of most dwarf irregular
galaxies are best described by a low level of star formation activity
which percolates across the gaseous disk.  However, large imaging studies
of dwarf irregular galaxies have shown that there are a few dIs
which have structural parameters similar to the BCD class \citep[e.g.,][]{PT96,vZ00}.  
These ``compact dIs'' have similarly small scale
lengths for their luminosity, but do not have extremely high star formation
rates at the present time.  Thus, these galaxies are likely to be examples of
post--burst BCDs.  Indeed, unlike the other dwarf irregular galaxies in the samples, the
``compact dIs'' have detectable color gradients from the inner to outer
regions of the stellar disk, similar to those found in samples of starbursting
dwarf galaxies \citep[e.g.,][]{MHWS95,DCPST97,DCC99}.  

The kinematic data presented here provides additional evidence that BCDs do not 
represent a common phase in the star formation history of gas--rich low mass galaxies.
BCDs have steeper rotation curves and lower specific angular momenta than ``typical''
dwarf irregular galaxies \citep[Figures 12 and 13, see also][]{MSSK98}.  However, this is
not to say that there is no overlap between BCDs and dIs in this regard
as well.  In order to accentuate the differences between BCDs and dIs, 
we have focussed here on examples taken from the extremes of both populations (LSB and
HSB).  In fact, dIs embody a continua of properties, from the low surface
brightness regime to the high surface brightness regime, and with a wide range
of star formation rates and gas content.  Thus, the marked differences seen
in Figures \ref{fig:gas} and \ref{fig:angmom} should not be used to infer
that post--burst BCDs are morphologically and dynamically distinct from
general dI classification.  Instead, it appears
that post--burst BCDs can be found within the general dI classification,
but rather than being a necessary phase in dwarf galaxy evolution, starbursting
episodes are rare events.

\subsection{Dwarf Galaxy Evolution and Cosmology}

As the first structures to collapse in hierarchical galaxy formation models,
dwarf galaxies play a crucial role in galaxy formation.  However, the interplay between
structure formation, the onset of star formation activity, and the
subsequent evolution of baryonic matter is complex and not well understood
at the present time.  Most cosmological models treat the physics of the
dark matter particles in great detail, while the complex processes of star
formation and gas dynamics are applied as simplified ``recipes''  
\citep[e.g.,][]{WF91,KWG93,CASFNZ94,HCFN95,SP99,CLBF00}.
Unfortunately, this broad brush approach to star formation means that
all dwarf galaxies are treated in a similar manner, regardless of their
physical attributes; further, none of the current models have the spatial resolution
to track the gas distribution in individual galaxies, and thus they cannot
determine if a galaxy is likely to form stars in a large burst of star formation
(centrally concentrated gas distributions), or in a more quiescent and constant manner 
(low density gas distributions).  Ultimately, the lack of spatial resolution
and the poor treatment star formation processes means that current cosmological
models cannot differentiate between low surface brightness dwarf galaxies
and high surface brightness dwarfs.

As discussed in Section 4.2.2, low surface brightness dwarf galaxies and
high surface brightness (starbursting) dwarf galaxies probably represent the 
extreme ends of a continua of dwarf galaxy properties.  One of the critical parameters
which still needs to be determined is the relative number density of starbursting
dwarf galaxies to quiescent dwarfs.  In particular, BCDs have gained attention recently 
due to the identification of bursting systems at intermediate and high redshifts.  
\citet{FB98} proposed that very intense and short bursts
of star formation in low mass galaxies (similar to the present day
BCDs) could be responsible for the faint blue galaxy excess found in studies
of high redshift luminosity functions \citep[e.g.,][]{CSH91}. 
However, if BCDs represent the extreme members of the dwarf irregular class, and occur
only rarely, then only a small fraction of dwarf galaxies will go through a starburst
phase.  As a consequence, the number density of starbursting dwarf galaxies as a function
of redshift may be lower than previously thought.

The possibility that there may be fewer starbursting dwarf galaxies at high
redshift also has severe implications for galaxy evolution models.  Most cosmological
models assume that the net effect of any star formation activity in a dwarf galaxy
is the immediate injection of heavy elements into the intergalactic medium
via galactic winds and other ``feedback'' processes \citep[e.g.,][]{SLGV99,CLBF00,E00}.  
These models predict rapid enrichment of the intergalactic medium, as low mass
halos (dwarf galaxies) collapse in the early universe.
If the majority of dwarf galaxies do not go through a starburst phase, 
however, this enrichment process may be less efficient than currently assumed.   

\section{Summary and Conclusions}
\label{sec:conc}

In summary, we have presented the results of high spatial resolution 
HI synthesis observations of six BCD/dEs.  Our results and conclusions
are summarized below.

(1) All six of the BCDs in this sample are rotation dominated systems.  
The velocity fields and position--velocity diagrams clearly indicate that 
these galaxies have significant velocity gradients across the gaseous disks 
(Figure \ref{fig:pv}).

(2) As in previous studies of BCDs \citep[e.g.,][]{vSS98}, the neutral gas 
density is centrally concentrated (Figure \ref{fig:gas}), which presumably  
facilitates the high star formation rate in these compact galaxies.

(3) Starbursting dwarf galaxies appear to have lower specific angular
momenta than similar luminosity low surface brightness dwarf irregular 
galaxies (Figure \ref{fig:angmom}).  The intrinsic angular momenta differences
may explain both the compact nature of BCDs and their tendency to have starburst 
episodes.

(4) The observed gas dynamics places strong constraints on the evolutionary
fate of starbursting dwarf galaxies.  Based on their gas kinematics, BCDs cannot
evolve passively into dwarf elliptical galaxies.  In contrast, both BCDs and dIs
are rotationally supported systems, and thus it is likely that post--burst
BCDs can be found within the general dwarf irregular classification.

These observations place the first strong kinematic constraints on BCD evolutionary
scenarios.  The fact that the gaseous disks are supported by rotation,
even for these most ``dE--like'' of the BCD class,  implies that
starbursting dwarf galaxies  will not evolve passively into dwarf elliptical galaxies. 
However, since it is possible to transfer angular momentum via merging or interactions 
between galaxies, BCDs in high density regions could evolve into dwarf ellipticals via
a more catastrophic evolutionary scenario.  In fact, the formation of dwarf elliptical
galaxies via ram pressure stripping of either gas--rich dwarf irregular galaxies or 
star--bursting dwarf galaxies is further substantiated by the strong morphology--density 
relationship in low mass galaxies.  

One potential concern is that the kinematic constraints presented here are for
the gaseous disks of the starbursting dwarf galaxies, while the kinematic
data for dwarf elliptical galaxies is almost entirely derived from observations
of the stellar population.
If the stars and gas are kinematically de--coupled in BCDs, the angular momentum
arguments presented here are less compelling.  In this regard, it is important to note that
of the few dwarf ellipticals which have been mapped in HI, at least one, NGC 205,
has a detectable velocity gradient in the gas disk \citep{YL97} which is not
detected in the stellar motions \citep{BPN91}.  Thus, further observations of
the stellar kinematics of BCDs will be necessary to definitely exclude the
possibility that BCDs evolve passively into dwarf elliptical galaxies when the 
starburst fades.

Nonetheless, the observed dynamics of BCDs appear to favor the hypothesis that post--burst
BCDs should be found within the large, diverse, dwarf irregular classification.
The definitive test of the BCD to dI evolutionary scenario will be to find examples
of post--burst BCDs which have been classified as dwarf irregular galaxies.
Studies of the structural parameters of dwarf irregular galaxies suggest that
the majority of dIs are not related to the BCD class, but that a small percentage
have similar morphology and color gradients \citep{vZ00,vZ01}.  If these compact
dIs are post--burst BCDs, they should also have similar gas distributions and gas
dynamics as the BCDs.  Thus, future HI synthesis observations of compact dIs should provide
a direct test of the BCD to dI evolutionary scenario, and definitively
determine if BCDs fade gracefully into more quiescent dIs, or if 
an altogether different, heretofore unknown, fate lies ahead.

\acknowledgements
We acknowledge the financial 
support by NSF grant AST95--53020 to JJS.  Partial support from
a NASA LTSARP grant No. NAGW--3189 (EDS) is gratefully acknowledged.
This research has made use of the NASA/IPAC Extragalactic Database (NED)   
which is operated by the Jet Propulsion Laboratory, California Institute   
of Technology, under contract with the National Aeronautics and Space      
Administration.

\begin{table}
\dummytable\label{tab:global}
%this table contains the list of global parameters
\end{table}

\begin{table}
\dummytable\label{tab:vlaobs}
%this table contains the list of configurations and obs. parameters
\end{table}

\begin{table}
\dummytable\label{tab:maps}
%this table contains the list of imagr parameters
\end{table}

\begin{table}
\dummytable\label{tab:others}
%this table contains the list of HI parameters of EXG 0123-0040 and EXG 2323+1816
\end{table}

\begin{table}
\dummytable\label{tab:colden}
%this table contains the list of peak column densities
\end{table}

\begin{table}
\dummytable\label{tab:hiparms}
%this table contains the list of HI parameters
\end{table}

\psfig{figure=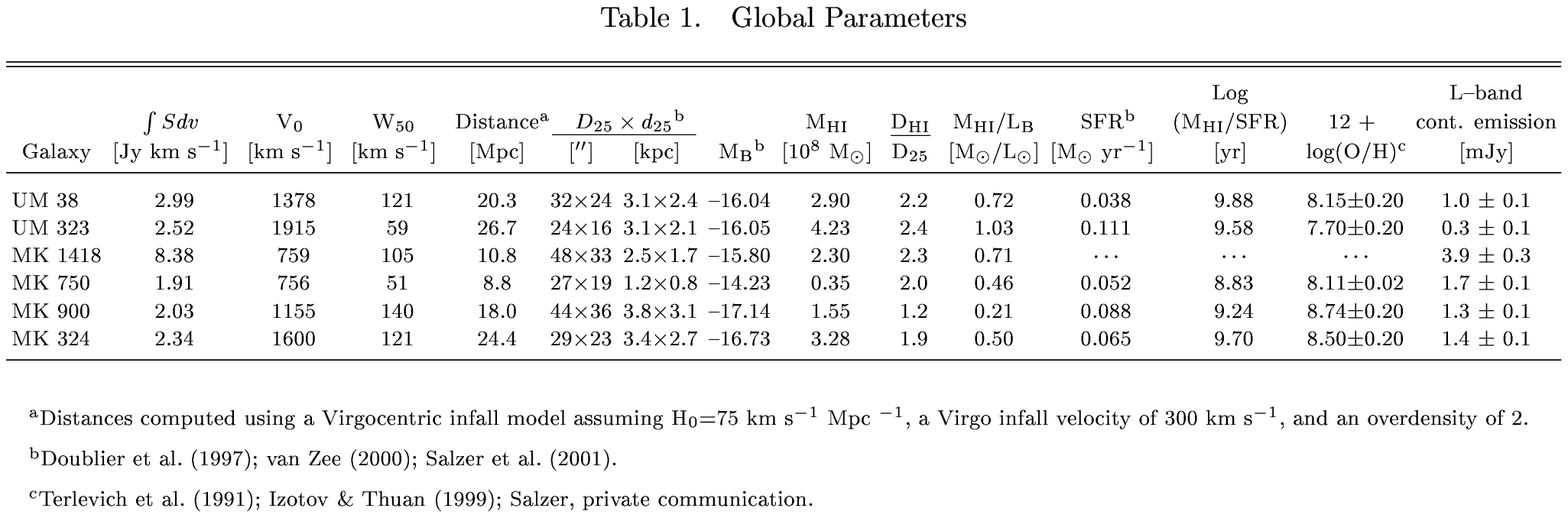,height=25.cm}

\psfig{figure=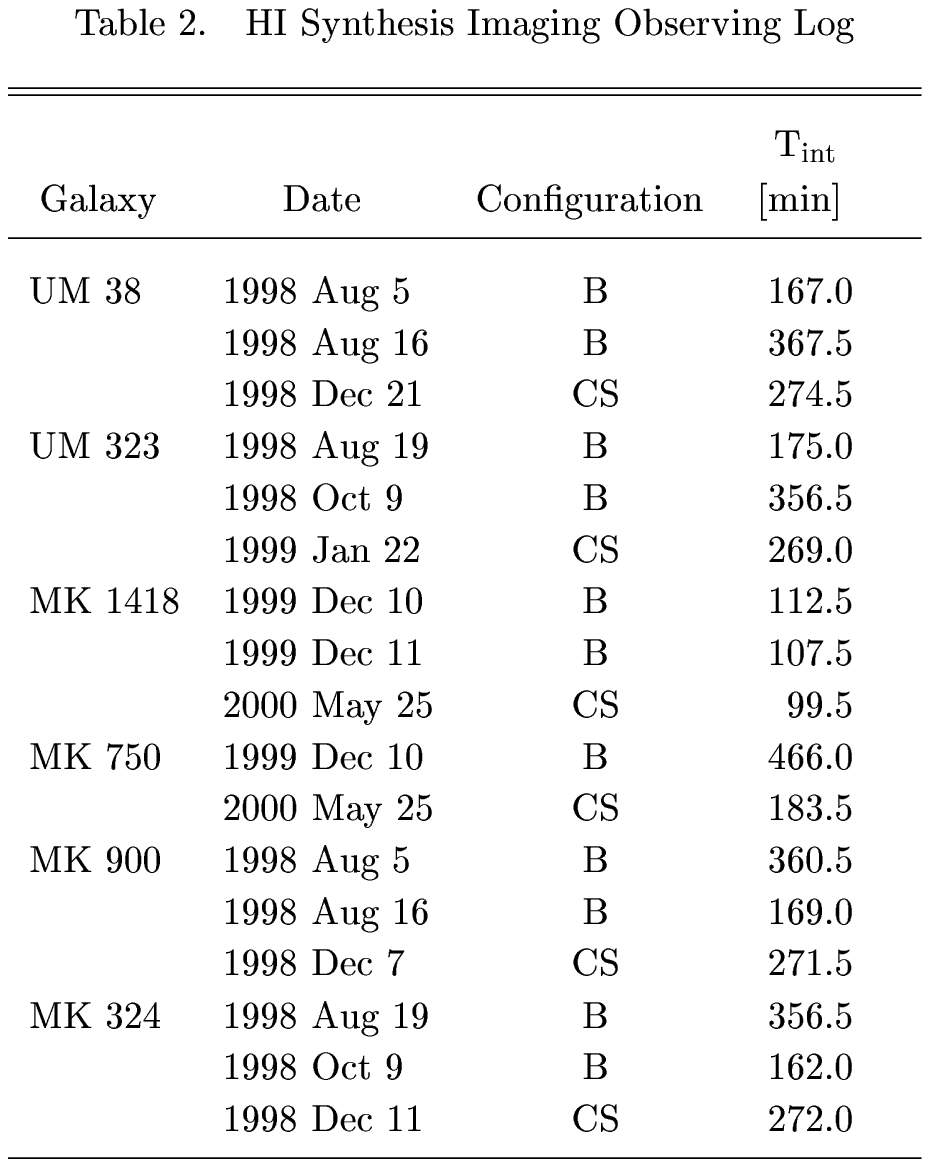,height=25.cm}

\psfig{figure=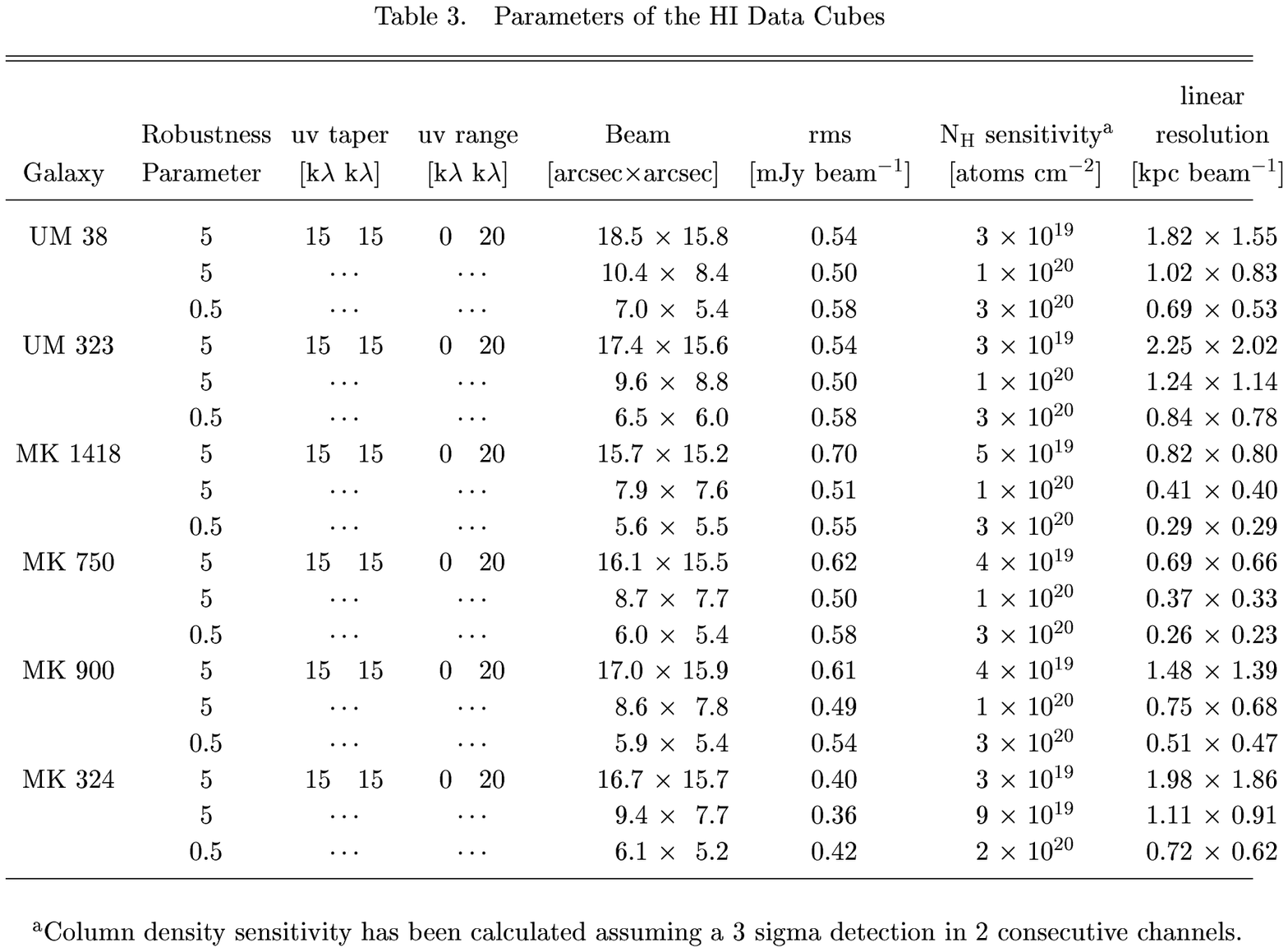,height=25.cm}

\psfig{figure=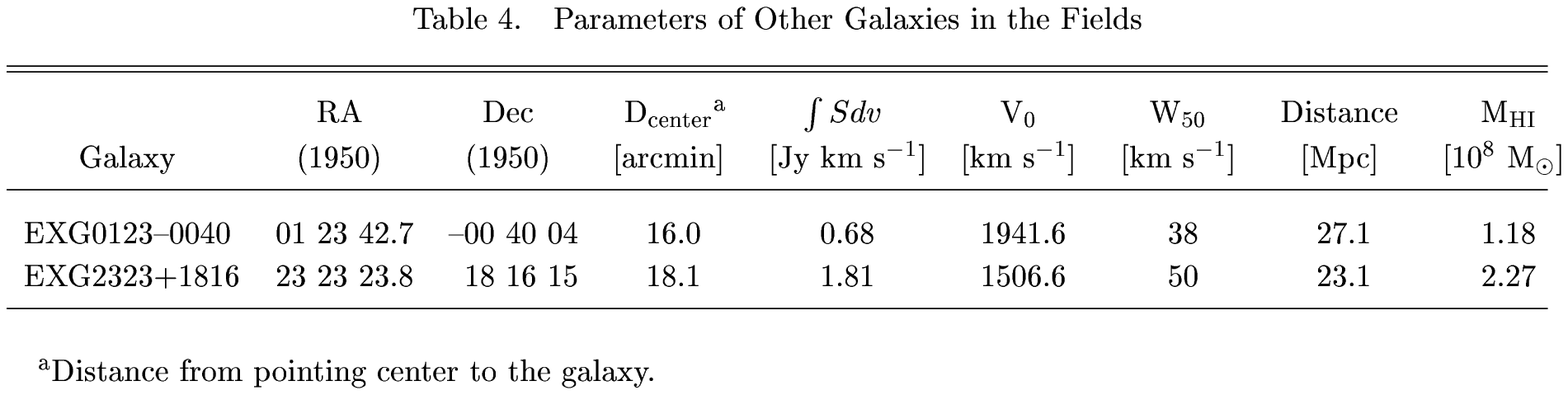,height=25.cm}

\psfig{figure=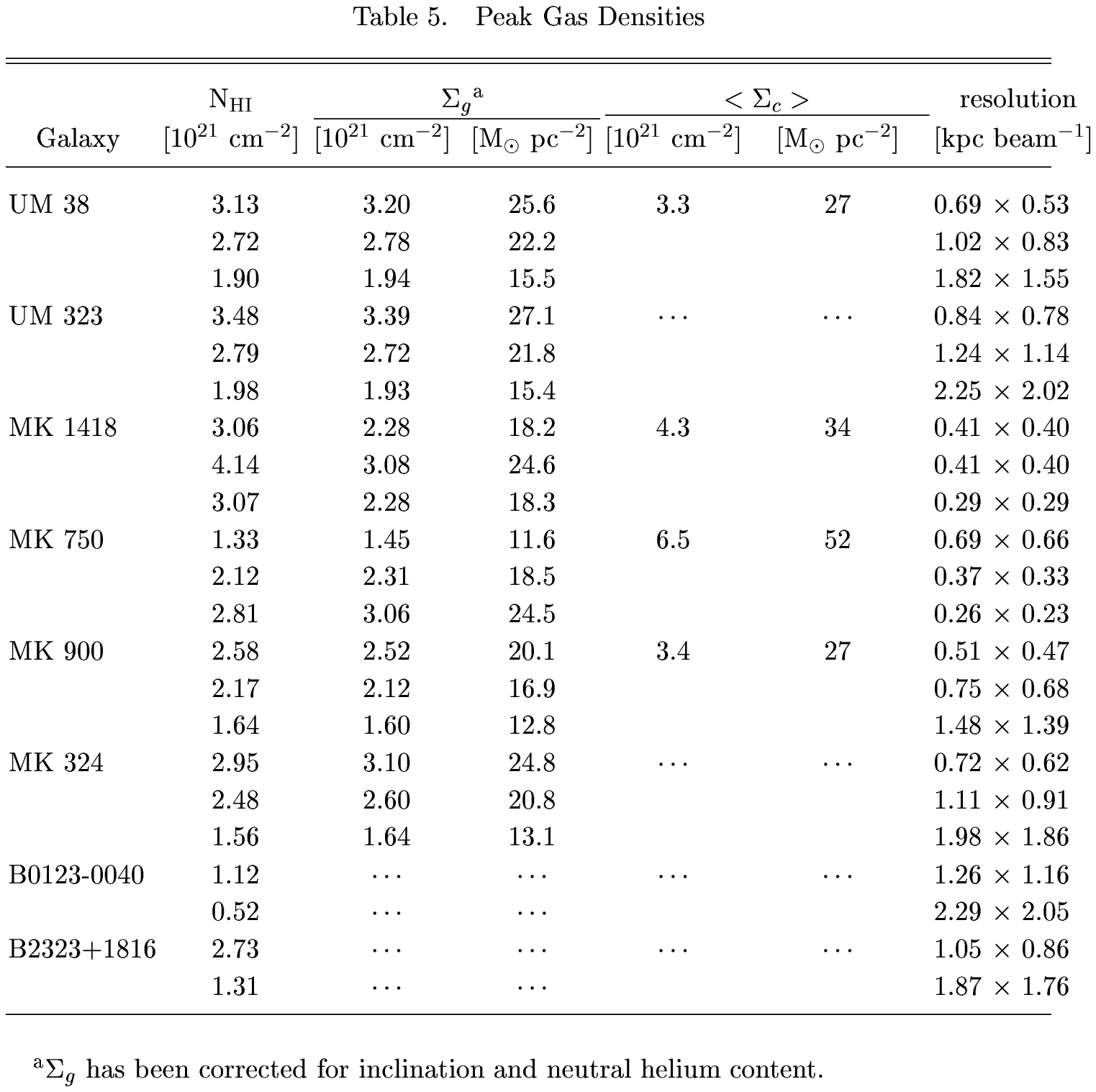,height=25.cm}

\psfig{figure=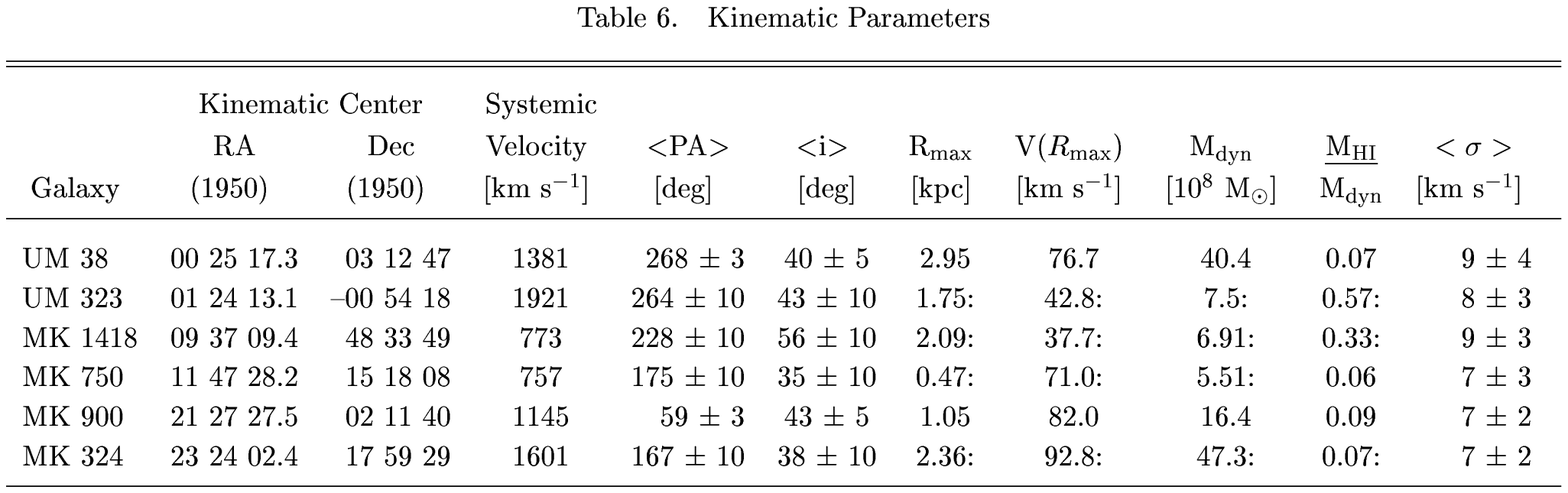,height=25.cm}

\end{document}